\documentclass[8.5pt,twoside,twocolumn]{article}
\usepackage[super,sort&compress,comma]{natbib} 
\usepackage{mhchem}
\usepackage{times,mathptmx}
\usepackage{times}
\usepackage{sectsty}
\usepackage{balance} 

\usepackage{amssymb} 

\usepackage{graphicx} 
\usepackage{lastpage}
\usepackage[format=plain,justification=raggedright,singlelinecheck=false,font=small,labelfont=bf,labelsep=space]{caption} 

\begin{document}

%
%
%
\twocolumn[
\begin{@twocolumnfalse}
\noindent\centering\LARGE{\textbf{Topological Defects, Surface Geometry and Cohesive Energy of Twisted Filament Bundles}}
\vspace{0.6cm}

\noindent\centerline{\large{Isaac R. Bruss, and Gregory M. Grason}}
\noindent\centerline{\small{\textit{Department of Polymer Science and Engineering, University of Massachusetts, Amherst, MA, 01003, USA}}}
\noindent\centerline{\textit{\small{\textbf{\newline}
(Dated: July 1st, 2013)}}}
%
\noindent \textbf{\small{\newline}}

%

\noindent \normalsize{
Cohesive assemblies of filaments are a common structural motif found in diverse contexts, ranging from biological materials such as fibrous proteins, to artificial materials such as carbon nanotube ropes and micropatterned filament arrays. In this paper, we analyze the complex dependence of cohesive energy on {\it twist}, a key structural parameter of both self-assembled and fabricated filament bundles. Based on the analysis of simulated ground states of cohesive bundles, we show that the non-linear influence of twist derives from two distinct geometric features of twisted bundles: (i) the geometrical frustration of inter-filament packing in the bundle cross-section; and (ii) the evolution of the surface geometry of bundles with twist, which dictates the cohesive cost of non-contacting filaments at the surface. Packing frustration in the bundle core gives rise to the appearance of a universal sequence of topological defects, excess 5-fold disclinations, with increasing twist, while the evolution of filament contact at the surface of the bundle generically favors twisted geometries for sufficiently long filaments. Our analysis of both continuum and discrete models of filament bundles shows that, even in the absence of external torque or intrinsic chirality, cohesive energy universally favors {\it twisted} ground states above a critical (length/radius) aspect ratio and below a critical filament stiffness threshold.}
\vspace{0.5cm}
 \end{@twocolumnfalse}]

\section{Introduction}
The superior mechanics afforded by the {\it twisting} of ropes and fibers has apparently been known to humans for tens of thousands of years \cite{1994, Kvavadze2009a}. In ``staple" yarns and fibers, spun from finite-length filaments, twist provides a geometric mechanism of cohesion, transmitting tensile strain along the fibers length to in-plane compression of the cross-section, locking the structure together via friction \cite{hearle1969structural}. In continuous strand assemblies, like wire ropes and cables, the helical ``pre-twist" geometrically relaxes the differential stretching/compression under bending, conferring a greater flexibility to the structure \cite{costello1997theory}.

Given the robust mechanical properties of twisted filament ropes and bundles, it is perhaps unsurprising that Nature incorporates this design motif into a host of biological structures at the molecular scale. Helically wound cellulose microfibrils provide mechanical reinforcement for walls of wood cells \cite{Fratzl2003a}, while helically twisted fibrils of extra-cellular protein filaments like collagen \cite{Hulmes1995} and fibrin \cite{parry2005fibrous} play crucial mechanical roles in animal tissue. Since molecular fibers and bundles of biofilaments are not assembled by hand, the spontaneous twist derives from the process of assembly itself. Molecular scale chirality of biomolecules---the fact that biofilaments are universally helical in structure---is broadly implicated as the driving force for handed inter-filament packing in twisted bundles \cite{Grason2007, Turner2003, Weisel01121987, Bouligand1985, Bouligand2008}, the filamentary analogues of the cholesteric and double-twisted textures of chiral liquid crystals \cite{Kamien1995, p1995physics}.

While the influence of the twisted geometry of ropes and fibers on mechanical properties---high strength under tension and extreme compliance to bending---is relatively well developed in the textile engineering literature, the question of how twist influences inter-filament contact, and therefore the cohesive energy in filament bundles, remains poorly understood \cite{Morimoto2012, Neukirch2003, Olsen2009}. In this article, we analyze the dependence of packing geometry and inter-filament cohesion on the twist of filament bundles based on a simple model of inter-filament, attractive interactions. The dependence of cohesive energy on twist is particularly relevant for bundles of molecular (nano-)scale filaments, such as carbon nanotubes or filamentous proteins, where thermodynamic contributions from inter-filament attractions may be significant compared to the mechanical costs of filament deformation. Even in the simplest model of filament cohesion that we analyze here, the geometry of twist has a significant influence on both the structure and energetics of minimal-energy filament bundles, owing to the surprisingly subtle nature of contact between filamentary objects.

\begin{figure*}
\centering
\includegraphics[width=17.5cm]{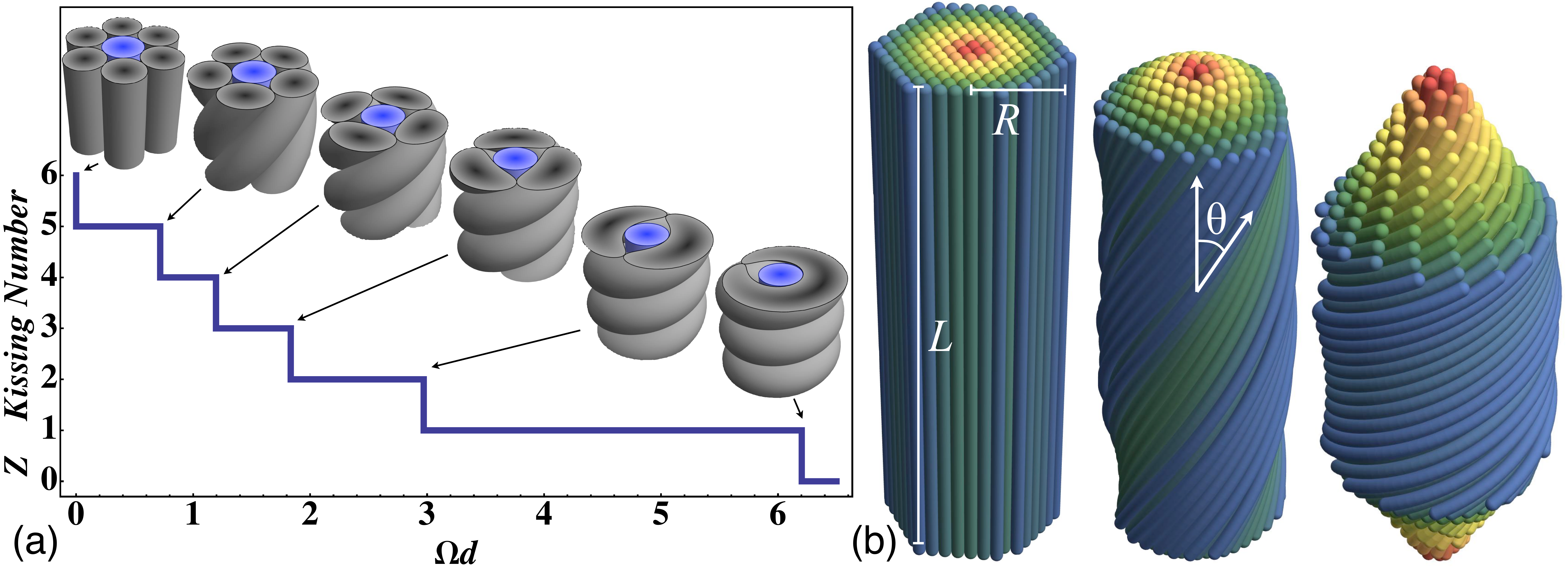}
\caption{(a) Kissing number, $Z$, for the central (blue) filament, vs rate of twist, $\Omega d$, where $d$ is the filament diameter. Example show the structures with the largest $\Omega d$ for a given $Z$. (b) Examples of three cohesive bundles comprised of 145 filaments of a fixed length, $L = 40 d$. From left to right, $\theta = 0^{\circ}$,$31.2^{\circ}$, and $61.2^{\circ}$. Additional parameters are shown: the filament contour length, $L$, the bundle radius, $R$, and the twist angle, $\theta$. Coloring is simply used to highlight radial depth of filament position.}
\label{fig:KissingNumber}
\end{figure*}

As we show in this study, the complex and non-linear twist-dependence of cohesive energy in multi-filament bundles derives from the interplay from two competing geometrical effects: i) {\it geometric frustration} of the bulk packing and ii) {surface energy} of non-contacting filaments at the boundary. The geometric frustration within the bulk may be visualized in terms of {\it local constraints} of packing in twisted filament bundles. A measure of these constraints is the {\it kissing number} \cite{Conway1998}, $Z$, of the central filament in a twisted bundle shown in Fig.\ \ref{fig:KissingNumber}a, which counts the maximum number of non-overlapping filaments in contact with the central filament.
The helical shape of the finite-diameter filaments surrounding the central filament implies that contacting neighbors occupy greater than $2 \pi/6$ of the planar angle surrounding the central (straight) filament. This means that at the center of the bundle, twist obstructs the six-fold hexagonal close-packing that is possible for straight and parallel fibers. Therefore $Z< 6$ for any finite degree of twist, measured by a non-zero {\it rate of twist}, $\Omega$, with $2 \pi/ \Omega$ defined as the helical pitch of the bundle. The impossibility of perfect six-fold lattice packing at the core has recently been shown \cite{Bruss2012} to derive from an exact mapping of the contact geometry of filaments in a twisted bundle on to the problem of packing of discs on a spherically-curved surface, whose Gaussian curvature is proportional to $\Omega^{2}$. This mapping implies that, like the ordering of particles on spherically-curved surfaces \cite{Bowick2009}, 1) the geometric frustration of filament spacing introduces inter-filament strains whose cost grows with bundle twist, and 2) in sufficiently large and twisted bundles, an excess of certain topological defects, 5-fold disclinations, are necessary components of the minimal-energy lattice packing \cite{Grason2012, Grason2010}. Hence, a quantitive analysis of the cohesive energy of twisted filament bundles requires careful accounting of the number and organization lattice defects, whose presence alters the inter-filament spacing throughout the bundle.

While the geometric frustration of inter-filament packing is largely localized near the interior core of twisted bundles, the energy cost of non-contacting filaments is distributed at the outer surface of the bundle, whose geometry is also strongly altered by twist. In untwisted bundles of uniform length filaments, the surface energy derives only from the cylindrical sides of the bundle, as the ends expose no length of non-contacting filaments. As a bundle twists, the constant length and helical shape of filaments implies that the vertical height of the bundle shortens with radial distance from the bundle center, resulting in a ``tapering" of the bundle profile that increases with $\Omega$. As the occupied volume within the bundle is largely conserved, the tapered bundle must also expand radially, leading to a complex progression of surface profiles with increasing twist, as shown in Fig.\ \ref{fig:KissingNumber}b. Due to this complex boundary shape, determining whether the surface energy of the bundle increases or decreases with twist becomes a delicate affair. The ends of twisted bundles expose non-contacting filament length, whereas the decreased area of the cylindrical sides reduce this exposure. The overall balance of surface energy depends critically on just three length parameters of the bundle: the filament contour length, $L$, the (untwisted) bundle radius, $R_0$, and the twist angle, $\theta$, which is the angle between the outermost filaments with the vertical twist axis.

In this article, we analyze the thermodynamic consequences of both packing frustration and surface geometry in twisted bundles based on a combination of numerical and analytical studies of a simple model of short-ranged cohesive interactions between finite-length filaments. In ref.\ \cite{Bruss2012}, we introduced and explored a simple model of cohesive interactions in $N$-filament twisted bundles to study the dependence of the minimal-energy filament packing on twist. Here, we extend the analysis of this model, by studying the evolution of the cohesive energy in with twist, accounting specifically for changes associated with filaments at the sides and ends of finite-length filament bundles. We show that this twist-dependence can be decomposed into two contributions: 1) a bulk, elastic contribution that accounts for the effects of frustration of inter-filament spacing in the bulk, and the subsequent lattice defects which relax frustration; and 2) a surface contribution, accounting for the total length of non-contacting filaments in the bundle. We develop a continuum theory of the surface energy, and show that the decrease of surface area at the bundle sides by twist dominates the cost of increased filament exposure at the ends, provided a large enough aspect ratio $L/R_0$. Importantly, this analysis shows that the relaxation of surface energy generically dominates the mechanical cost of twist-induced filament bending, provided that inter-filament cohesion (per unit contact length) exceeds a universal critical value of $\gamma_{0} \gtrsim 1.297 B/R_{0}d$, where $B$ is the bend modulus of filaments and $d$ is the filament diameter. Based on these results and the results of our numerical model, we are led to the surprising conclusion that---even in the absence of intrinsically chiral interactions driving twist---sufficiently long and flexible filaments {\it generically favor a twisted geometry}, provided that bundle packing adopts the distribution of lattice defects required to mitigate the frustration at the core.

The remainder of this article is organized as follows. In section \ref{sec:DiscreteModel} we introduce the discrete model cohesive interactions in $N$-filament, twisted bundles. In section \ref{sec:CoreEnergy} we analyze the structure and energetics of ground-state packings of bundles in the limit of infinite length filaments. In section \ref{sec:SurfEnergy}, we develop a continuum analysis of the surface energy in finite-length bundles, which we compare to the infinite-length results. Additionally, we analyze the thermodynamics of surface energy in finite-length twisted bundles in combination with the mechanical costs of filament bending. In section \ref{sec:DiscreteFiniteL}, we consider the energetics of our discrete model for the case of finite-length bundles, analyzing the thermodynamics while accounting for both bulk and surface packing of filaments, and determine the conditions for which minimal-energy bundles have non-zero twist. We conclude in section \ref{sec:Discussion} with a discussion of the importance of filament packings on the cohesive energy, as well as an analysis of the predictions of our model in the context of a range of cohesive filament systems.

\section{Discrete model of filament cohesion in twisted bundles}
\label{sec:DiscreteModel}
In this section we introduce a model of $N$-filament twisted bundles possessing short-range cohesive interactions. First, we derive the form of the pair-wise contact interactions between filaments in terms of the local geometry of inter-filament contact. We then analyze the specific geometric effects on contact of filaments within the bounds of a twisted bundle geometry.

\subsection{Filament contact and cohesive interactions}
\label{sec:FilCohesion}
Here, we summarize the relationship between pair-wise interactions between portions of filaments (arc-length elements) and the cohesive energy expressed in terms of the {\it local geometry} of inter-filament contact. In this study, we consider bundles of homogenous and mutually-attractive filaments, where the interactions and mechanics can be described purely in terms of the shape of the filament center line, ${\bf X}_i (s_i)$, which describes the position of filament $i$ at arc-position $s_i$ along backbone. We assume that the filament pairs interact via the summation of short-ranged, pairwise interactions between arc-length elements, such that the interaction between filaments $i$ and $j$ has the form,
\begin{equation}
E_{ij} = \int ds_i \int ds_j V \big(|{\bf X}_i (s_i) - {\bf X}_j (s_j)|\big) ,
\end{equation}
where $V(r)$ is an isotropic, finite-range potential describing interactions between length elements separated by a distance $r$.

While a given length element at $s_i$ on the filament $i$ interacts with the entire length on $j$, the finite range of segment interactions generically implies that interactions of filament $i$ at $s_i$ are dominated by region of filament $j$ closest to ${\bf X}_i(s_i)$, which we call the {\it contact region}. Sufficiently, far from the filament ends, we may describe the contact geometry of $s_i$ with filament $j$ straightforwardly, in terms of $s^*_j(s_i)$, a function that maps the arc-position on $i$ to the position on $j$ closest to ${\bf X}_i(s_i)$, which we call the {\it point of contact}. Notably, this allows for a generic expansion of the shape of the contacting filament around the point of contact, in terms of the local geometry of $j$ and distance along $j$ from this point, $\delta s_j = s_j- s^*_j(s_i)$,
\begin{equation}
{\bf X}_j (s_j) = {\bf X}_j (s^*_j ) + \delta s_j {\bf T}_j + (\delta s_j)^2 \frac{ \kappa_j }{2} {\bf N}_j + O\big[ (\delta s_j)^3\big] ,
\label{eq:XjExpansion}
\end{equation}
where ${\bf T}_j$, ${\bf N}_j$ and $\kappa_j$, are the tangent, normal and curvature of filament $j$ at $ s^*_j(s_i)$ ~\cite{Kamien2002}. From this expression we find the interfilament distance, $\pmb{\Delta} \equiv {\bf X}_j (s_j) - {\bf X}_i (s_i)$, from
\begin{equation}
\label{delta2}
|\Delta(\delta s_j)|^2= |\Delta_{ij} |^2 + (\delta s_j)^2 \big( 1+ \kappa_j\pmb{\Delta}_{ij} \cdot {\bf N}_j \big) + O\big[ (\delta s_j)^3\big],
\end{equation}
where $\pmb{\Delta}_{ij} = {\bf X}_j (s^*_j) - {\bf X}_i (s_i)$ is the {\it distance of closest contact} from $s_i$ to filament $j$, such that $\pmb{\Delta}_{ij}  \cdot {\bf T}_j =0 $. These parameters are shown schematically in Fig.\ \ref{fig:FilPara}.

Assuming that $V(r)$ is sufficiently short-ranged compared to filament length and curvature, we may use eqn (\ref{delta2}) to derive the cohesive energy contribution of the length element at $s_i$,
\begin{equation}
\label{contactenergy}
 d E_{ij} = ds_i\int_{-\infty}^{+\infty} d(\delta s_j) V\big(|\Delta(\delta s_j)| \big) = \frac{ \gamma(\Delta_{ij}) }{\sqrt{ 1+ \kappa_j \pmb{\Delta}_{ij} \cdot {\bf N}_j } } ds_i, 
\end{equation}
where
\begin{equation}
\gamma(\Delta_{ij}) = \int_{- \infty}^{+\infty} du~ V \Big( \sqrt{\Delta_{ij}^2 + u^2 }\Big) ,
\end{equation}
is a local {\it contact potential}, describing the cohesive energy per unit length of filament $i$ in contact with filament $j$. This potential is a function of the local distance of closest contact between the filaments. The derivation of this local inter-filament cohesive energy is fully general for any short-ranged potential. In the remainder of the article we study the case of a Lennard-Jones interaction between length elements, $V(\Delta) = \epsilon\big[ (\sigma/\Delta)^{12}- 2 (\sigma/\Delta)^6 \big]$, for which the contact potential becomes,
\begin{equation}
\label{gamma}
\gamma (\Delta_{ij}) =\gamma_{0} \bigg[ \frac{5}{6} \Big(\frac{d}{\Delta_{ij}} \Big)^{11} - \frac{11}{6} \Big(\frac{d}{\Delta_{ij}} \Big)^5 \bigg],
\end{equation}
which like the Lennard-Jones, has stiff repulsion at short-range and soft attraction at long range. This potential is characterized by an attractive well of depth $\gamma_{0} = 1.6862 \epsilon\sigma$, at a preferred spacing $d = 0.9471 \sigma$, which we denote as the filament diameter. 

Eqn (\ref{contactenergy}) describes how short-ranged, pair-wise interactions between all length elements of a filament pair can be formulated in terms of functions of the local contact geometry (e.g.\ $\pmb{\Delta}_{ij}$, $\kappa_j$, and ${\bf N}_j$) integrated over the contacting length of a single filament (in this case, filament $i$). In the following section, we derive the contact geometry for filament pairs in twisted bundles.

\subsection{Filament contact geometry in twisted bundles}
\label{sec:FilTwisted}
\begin{figure}[h]
\centering
\includegraphics[width=8cm]{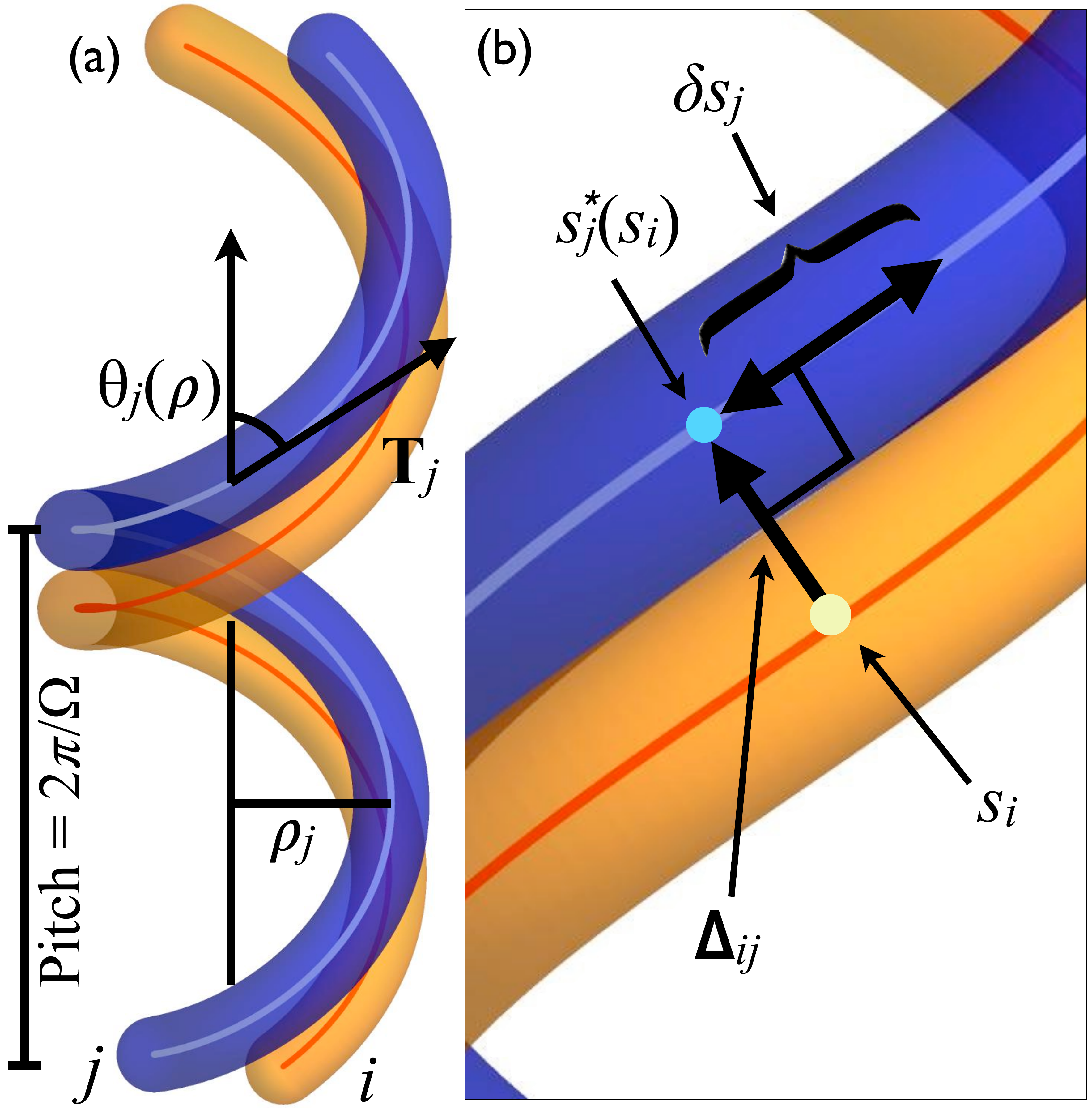}
\caption{(a) One and a half pitch lengths of a contacting helical filament pair, $i$ and $j$. The helical radius and helical angle for filament $j$ are $\rho_{j}$ and $\theta_{j}(\rho)$ respectively. (b) A view of a region of contact, showing the filament parameters defined in eqns (\ref{eq:XjExpansion}) and (\ref{delta2}).}
\label{fig:FilPara}
\end{figure}

In this section, we derive the geometry of inter-filament contact in bundles twisted around the central $\hat{z}$ axis at a uniform rate of rotation $\Omega$, which has been presented in ref.\ \cite{Bruss2012}. We consider bundles whose packing is {\it homogeneous} along their length, such that cross-sectional packing in a plane perpendicular to $\hat{z}$ differs only by the rigid rotation about the $\hat{z}$ axis. Defining the position of filament $i$ at a common plane $z=0$ (arc-position $s_i=0$), by the polar coordinates $(\rho_i,\phi_i)$, the shape of the filament backbone follows the helical curve,
\begin{equation}
\label{singlefil}
{\bf X}_i (z_i) = \rho_i \cos \big( \phi_i + \Omega z_i \big) \hat{x} +\rho_i \sin \big( \phi_i + \Omega z_i \big) \hat{y} + z_i\hat{z},
\end{equation}
where it is convenient to express position in terms of vertical height $z_i =s_i \cos \theta (\rho_i)$, where $ \theta (\rho) = \arctan (\Omega \rho)$ is the helical angle of filament $i$ with respect to $\hat{z}$, shown in Fig.\ \ref{fig:FilPara}a~\footnote{Note, that $\theta(\rho)$ refers to the {\it local} tilt angle of filaments at a radius $\rho$ in the bundle, where as, in our notation, when tilt angle appears without an explicit radius, as $\theta$, it refers to helical angle at the outer radius of the bundle, which we call the {\it twist angle} of the bundle.}.
The unit normal and curvature are easily calculated from the second-derivative of ${\bf X}_j$ with respect to arc-length,
\begin{align}
&\kappa_j = \frac{ \Omega^2 \rho_j}{1+(\Omega \rho_j)^2 } \\
&{\bf N}_j = - \cos \big( \phi_j + \Omega z_j \big) \hat{x} -\sin \big( \phi_j + \Omega z_j \big) \hat{y}.
\end{align}

For a pair of filaments, $i$ and $j$, it is convenient to express the separation between curves in terms of a vertical offset, $z_{ij} = z_i -z_j$, and the angular separation in the plane, $\phi_{ij} = \phi_i - \phi_j$,
\begin{equation}
\label{delsq}
\Delta^2(z_{ij}) = \rho_i^2+ \rho_j^2-2 \rho_i \rho_j \cos \big( \phi_{ij} + \Omega z_{ij} \big)+ z^2_{ij}.
\end{equation}
The distance of closest contact between $i$ and $j$ is determined by the minimization of $\Delta^2(z_{ij})$, with respect to the vertical offset between contacting points, to find the height separation at the distance of closest contact, $z_{ij}^*$. From $d \Delta^2(z_{ij})/d z_{ij}=0$ we find a transcendental equation satisfied by $z_{ij}^*$
\begin{equation}
\label{zstar}
\Omega z_{ij}^* = - \Omega^2 \rho_i \rho_j \sin \big( \phi_{ij} +\Omega z_{ij}^* \big) .
\end{equation}
To make practical use of this condition, we examine the solutions in the limiting cases where $\Omega^2 \rho_i \rho_j$ is either small or asymptotically large. The former case corresponds to either small twist, or filament positions sufficiently close to the center of the bundle, in which case the Taylor series expansion of eqn (\ref{zstar}) yields,
\begin{equation}
\lim_{\Omega^2 \rho_i \rho_j \ll 1 } \Omega z_{ij}^* = - \frac{ \Omega^2 \rho_i \rho_j \sin \phi_{ij} }{1+\Omega^2 \rho_i \rho_j \cos \phi_{ij} } + O\big[ (\Omega^2 \rho_i \rho_j)^3 \big],
\end{equation}
which represents a modest tipping out-of-plane of the distance of closest contact between azimuthally-separated filaments. In the opposite limit, filaments are far from the core of the bundle in comparison to their helical pitch. Since $z_{ij}^*$ is always strictly less than the pitch, the left-hand side of eqn (\ref{zstar}) is never larger than $2 \pi$ in magnitude, hence, in the limit $\Omega^2 \rho_i \rho_j \gg 1$, the solution for $z_{ij}$ must satisfy $\sin ( \phi_{ij} + z^*_{ij} ) \propto (\Omega^2 \rho_i \rho_j)$ so that in the asymptotic limit of large helical angles, we have
\begin{equation}
\lim_{\Omega^2 \rho_i \rho_j \gg 1 } \Omega z_{ij}^* = - \phi_{ij} ,
\end{equation}
which shows that the inter-filament distance of closest contact is nearly vertical far from the core of bundle.

These simple results can be incorporated into an approximate formula for vertical contact separation that interpolates between the two asymptotic limits,
\begin{equation}
\label{zstar_app}
\Omega z_{ij}^* \simeq - \arctan \Big( \frac{ \Omega^2 \rho_i \rho_j \sin \phi_{ij} } { 1+ \Omega^2 \rho_i \rho_j \cos \phi_{ij} } \Big) .
\end{equation}
We employ this form of $z^*_{ij}$ to approximate distance of closest contact in our numerical studies via the relation $\Delta_{ij} = \Delta(z^*_{ij})$. Notably, this approximation of $\Delta_{ij}$ becomes poor only for filament pairs distant from the center (large $\rho$) and on opposing sides of the bundle ($\phi_{ij} \approx \pi$), thus providing an accurate description for filament interactions in twisted bundles incorporating sufficiently short-ranged potentials ($|\Omega d| \ll 1$).

\section{Discrete model simulations: Core packing and energetics of twisted bundle ground states}
\label{sec:CoreEnergy}
In this section, we study the ground-state packings of twisted filament bundles by numerical simulation of our discrete model. Here, we expand upon the study of the evolution of filament packing with bundle twist presented in ref.\ \cite{Bruss2012}, correlating non-linear structural changes of the ground state with the complex twist dependence of the cohesive energy. As we find below, the appearance of lattice defects in the cross-sectional packing underlies certain universal features of the twist-dependence of energy. We first review our simulation method, and then present the results of simulations for defect structures of ground-state packings. We then proceed to analyze the generic form of the twist-dependent cohesive energy in our discrete model and its connection to the structural evolution of optimal packing.

\subsection{Numerical simulations of twisted bundle ground states}
\label{sec:NumSims}
To study the twist-dependence of the ground-state packing and cohesive energy of filament bundles, we numerically minimize the sum of pair-wise interactions described by eqns (\ref{contactenergy}) and (\ref{gamma}), based on the approximation of $\Delta_{ij}=\Delta(z_{ij}^*)$, described previously in eqn (\ref{zstar_app}). Since the interfilament contact geometry is constant along the bundle height, the cohesive energy for the $i$ and $j$ pair becomes,
\begin{equation}
\label{discreteEij}
 E_{ij}= \frac{ \gamma(\Delta_{ij})  L_{ij} }{\sqrt{ 1+ \kappa_j \pmb{\Delta}_{ij} \cdot {\bf N}_j } },
\end{equation}
where $L_{ij}$ is defined as the length of filament $i$ in contact with filament $j$. In this section, we consider only the case of infinite length, where filament contact is maintain along the full contour $L_{ij}=L$, provided that we take $i$ to be the ``outer" filament ($\rho_i\gtrsim \rho_j$). In this way we may neglect any explicit change of filament contact length at the {\it ends} of the bundle in the $L \to \infty$ limit.

We numerically optimize the total cohesive energy for bundles of fixed $N$ and $\Omega$. Our approach, described in detail in ref.\ \cite{Bruss2012}, generates $500-2000$ random initial configurations of in-plane filament positions, $\{{\bf X}_i (s_{i} = 0) \}$, and seeks a minimal-energy configuration of in-plane filament positions using the method of steepest descent. From this, we have found the minimum energy packings for bundles with $N$ between 16 and 196, and a multitude of twist angles between $0^{\circ}$ and $\sim80^{\circ}$~\footnote{The energy landscape of twisted bundles is particularly complex and possesses a large number of nearly degenerate minima. Given the finite number of random initial configurations samples for a give filament number and twist, our method is not guaranteed to resolve the exact ground state structure. Notwithstanding these computational challenges, our approach achieves bundle packings sufficiently close to the ground state, such that their structural evolution with twist is representative of the true ground-state structure of twisted bundles.}

\subsection{Defects in filament bundles}
\label{sec:DefectsInBundles}
The numerical simulations described above produce the ground-state positions and energies of filaments in the cross section of a bundle, given the input of $N$ and $\Omega$. These resultant packings are analyzed in terms of the network of {\it nearest-neighbor} contacts. The bond network is formally defined through a specialized Delaunay triangulation that takes into account the discrepancy in the separation between filament positions in the $xy$ plane, and the true distance of closest contact between filament pairs~\footnote{As described in ref.\ \cite{Bruss2012}, to determine the contact network, in-plane filament positions are transformed via a conformal mapping in the plane. This mapping has the property that (infinitesimally) small in-plane separations in the mapped coordinates along the {\it radial} and {\it azimuthal} directions correspond to the same interfilament distance. The nearest-neighbor bond network is computed by a Delaunay triangulation of these mapped coordinates that correctly account (locally) for the distances of closest contact rather than the in-plane distances.}. The triangulation of positions produces a unique network of non-overlapping bonds, whose ``nearest neighbor" connections serve as a measure of the local packing geometry. While in the case of an untwisted bundle the network universally describes the 6-fold, hexagonal packing, for sufficiently large twist, uniform 6-fold coordination is not maintained and the packing becomes interrupted by {\it topological defects}.

A primary type topological defect in 2D crystalline packings is a {\it disclination}, which describes the breakdown of long-range $n$-fold orientational symmetry in the lattice at a singular point \cite{nelson2002defects}. In a hexagonal crystal, low-energy disclinations are typically of two types: 5-fold and 7-fold. At the core of a 5-fold disclination, is a lattice site (corresponding to a vertex in the bond network) with five nearest neighbors, one fewer than the six neighbors of the perfect lattice. A disclination may be introduce in a hexagonal lattice by cutting out and removing a $60^{\circ}$ wedge, and stretching it back to close the gap. While the large elastic strains generated throughout the crystal generically make even single disclinations prohibitively expensive in most types of (planar) 2D crystals, our simulations reveal that disclinations are necessary components the minimal-energy packings of twisted filament bundles.

Following the topological characterization of disclinations in crystalline solids\cite{nelson2002defects}, we define the total topological charge of disclinations as
\begin{equation}
Q \equiv \sum_{n} (6-n)V (n),
\label{eq:DiscCharge}
\end{equation}
where $n$ is a coordination number of nearest neighbors belonging to a filament, and $V_{n}$ is the number of {\it internal} filaments that possess $n$ nearest neighbors. Non-internal, or {\it boundary} filaments, are distinguished by having at least one neighbor bond on the outer edge of the bond network.

\begin{figure*}
\centering
\includegraphics[width=17.5cm]{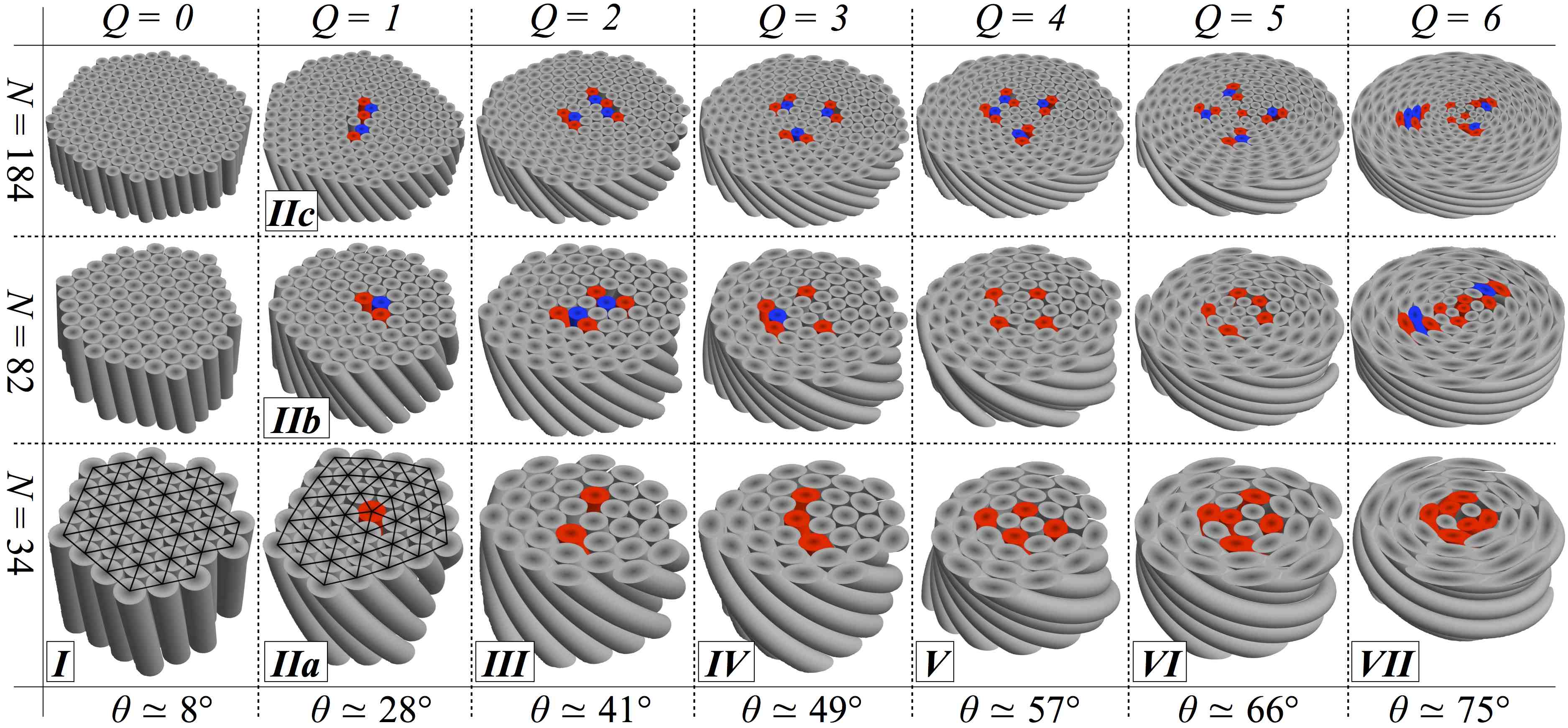}
\caption{Example cross sections of for three example bundle sizes and seven example twist angles, $\theta$, showing two trends: 1) the topological charge, $Q$, increases with twist angle, and 2) the total number of defects, $N_{Disclination}$, increases with bundle size. 5 and 7-fold disclinations are colored red and blue respectively. The bottom left two cross sections show their triangulated nearest neighbor bond network. The roman numeral labels match specific bundles to their locations within Fig.\ \ref{fig:SimPhaseD}.}.
\label{fig:BundleSpread}
\end{figure*}

Throughout this aritlcle, we consider the structure and energetics of three discrete bundle sizes in details: small ($N=34$); intermediate ($N=82$); and large ($N=184$). In Fig.\ \ref{fig:BundleSpread} we show the evolution of ground-state packings these three bundle sizes. The packing of each of these bundles exhibits a common trend with increasing twist. Packings evolve from defect-free $Q=0$ at zero twist, to an increasing value of topological charge---characterized a {\it universal excess 5-fold defects} with increasing twist. As the intermediate and large bundle packings illustrate, while the ground-state packing may include 7-fold disclinations, {\it negatively-charged} defects are always sufficiently outnumbered such that the net charge {\it increases} with twist.

\begin{figure*}
\centering
\includegraphics[width=17.5cm]{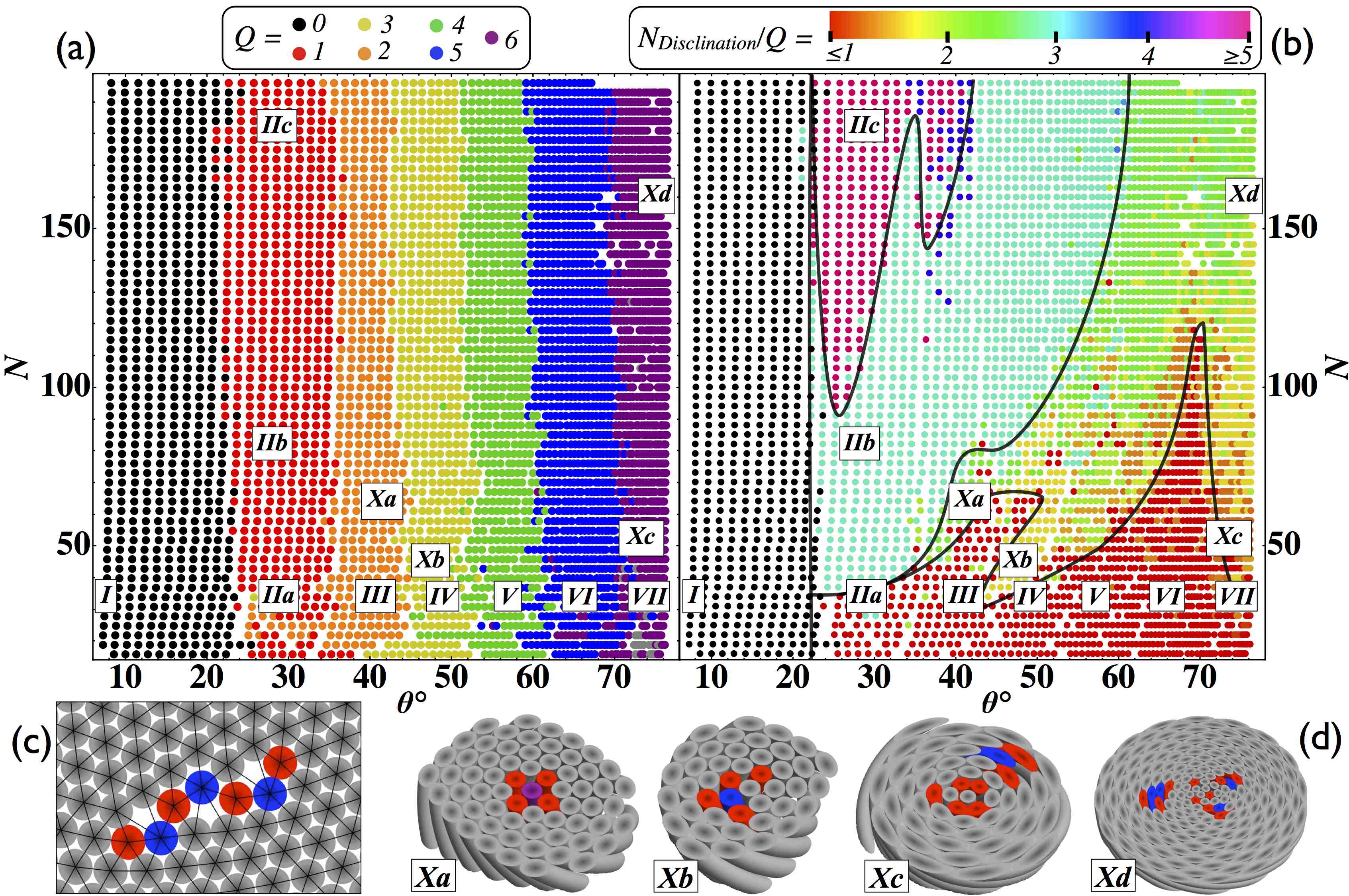}
\caption{(a) Phase diagram of net disclination charge, $Q$, with $\theta$ and $N$. (b) Phase diagram of $N_{Disclinations}/Q$, with twist angle $\theta$, and number of filaments, $N$. Black lines are shown to roughly delineate regions of qualitatively distinct ground-state packing. The roman numeral labels correspond to bundle cross sections seen in Fig.\ \ref{fig:BundleSpread} and part (d). (c) A zoomed-in view of a seven disclination long $Q=1$ scar, with bond network shown, for a bundle with $N=196$ and $\theta= 25.7^{\circ}$. In (d) we show five examples of bundles that exist in the variable-size defect region. 5, 7, and 8-fold disclinations are colored red, blue, and purple respectively. }
\label{fig:SimPhaseD}
\end{figure*}

By analyzing the topological charge, we construct a phase diagram bundle ground states in terms of twist angle, $\theta$, and number of filaments, $N$, shown if Fig.\ \ref{fig:SimPhaseD}a.
Notwithstanding some obvious limitation of small bundle sizes to reach large values of $Q$, it becomes clear that the optimal value of topological charge is predominantly determined by $\theta$, and largely independent of, $N$. $Q$ reaches a maximum of six at $\theta \gtrsim 70^{\circ}$. Further simulations show that $Q$ does not increased above 6, even for twist angles of up to $87^{\circ}$. We discuss the geometric origin of this universal dependence of $Q$ on $\theta$ in the next section.

While the net charge is largely fixed at a given twist angle, there are many ways to achieve a particular value of $Q$. Specifically, 5 and 7-fold disclinations that appear in ``neutral" pairs do not adjust this value. A commonly observed example occurs for $Q = 1$ bundles, which have $n$ 5-fold disclinations and $(n-1)$ 7-fold disclinations. For large bundles this feature becomes important, as shown in the 2nd column of Fig.\ \ref{fig:BundleSpread} (bundles $IIa$, $IIb$, and $IIC$). While these three examples have the same twist angles and hence maintain $Q = 1$, the total number of disclinations, $N_{Disclination}$, increases from 1 to 3 to 5 for small, intermediate and large bundles, respectively.

A second phase portrait of the number of disclinations, $N_{Disclination}$, per charge, $Q$, is show in Fig.\ \ref{fig:SimPhaseD}b. Unlike the net disclination charge, the optimal value of $N_{Disclination}/Q$ in ground-state bundles varies with both $N$, as well as $\theta$. In Fig.\ \ref{fig:SimPhaseD}b we have roughly delineated the $Q\neq0$ behavior into four regions. In the small-$N$ red region, $N_{Disclination} = Q$, and bundles contain primarily only 5-fold disclinations, without excess 5-7 pairs. For larger filament number, shown as a light blue region, $N_{Disclination} =  3 Q$. Ground states within this region contain structures such as $IIb$, where there are one 7-fold and two 5-fold disclinations per topological charge. At even larger bundle sizes is the pink region that continues this trend, i.e.\ there are two 7-fold and three 5-fold disclinations per topological charge, as seen in $IIc$. This trend continues with increasing $N$, reaching $N_{Disclination} = 7$ (as seen in Fig.\ \ref{fig:SimPhaseD}c) and higher for simulations not shown. A final region is loosely defined at the upper limits of twists, where packings identified in our ground-state search belong to a more complex taxonomy, possessing in general, non-integer values of $N_{Disclination}/Q$. This is achieved using solely 5 and 7-fold disclinations, like in $Xb$, $Xc$, and $Xd$, or with higher-charge disclinations like the 8-fold disclination shown in $Xa$ (Fig.\ \ref{fig:SimPhaseD}d).

\subsection{Mapping between twisted bundles and curved surfaces}
\label{sec:Mapping}
In the previous section, numerical simulations of twisted bundles revealed a complex structural evolution of ground-state filament packings with progressively increasing twist, marked by an increasing number of lattice defects in the cross-sectional order. The complex spectrum of defects is characterized by a universal dependence of the topological charge of the packing on the twist angle. In the next section, we show that transitions in the topological charge of the packing correspond to sharp, singular features in the twist-dependence of cohesive energy. As the twist-dependence of topological charge underlies a key feature of bundle energetics, in this section we briefly review the connection between twisted bundle packing and packing on spherically-curved surfaces, first described in the ref.\ \cite{Bruss2012}.

\begin{figure}[h]
\centering
\includegraphics[width=8.0cm]{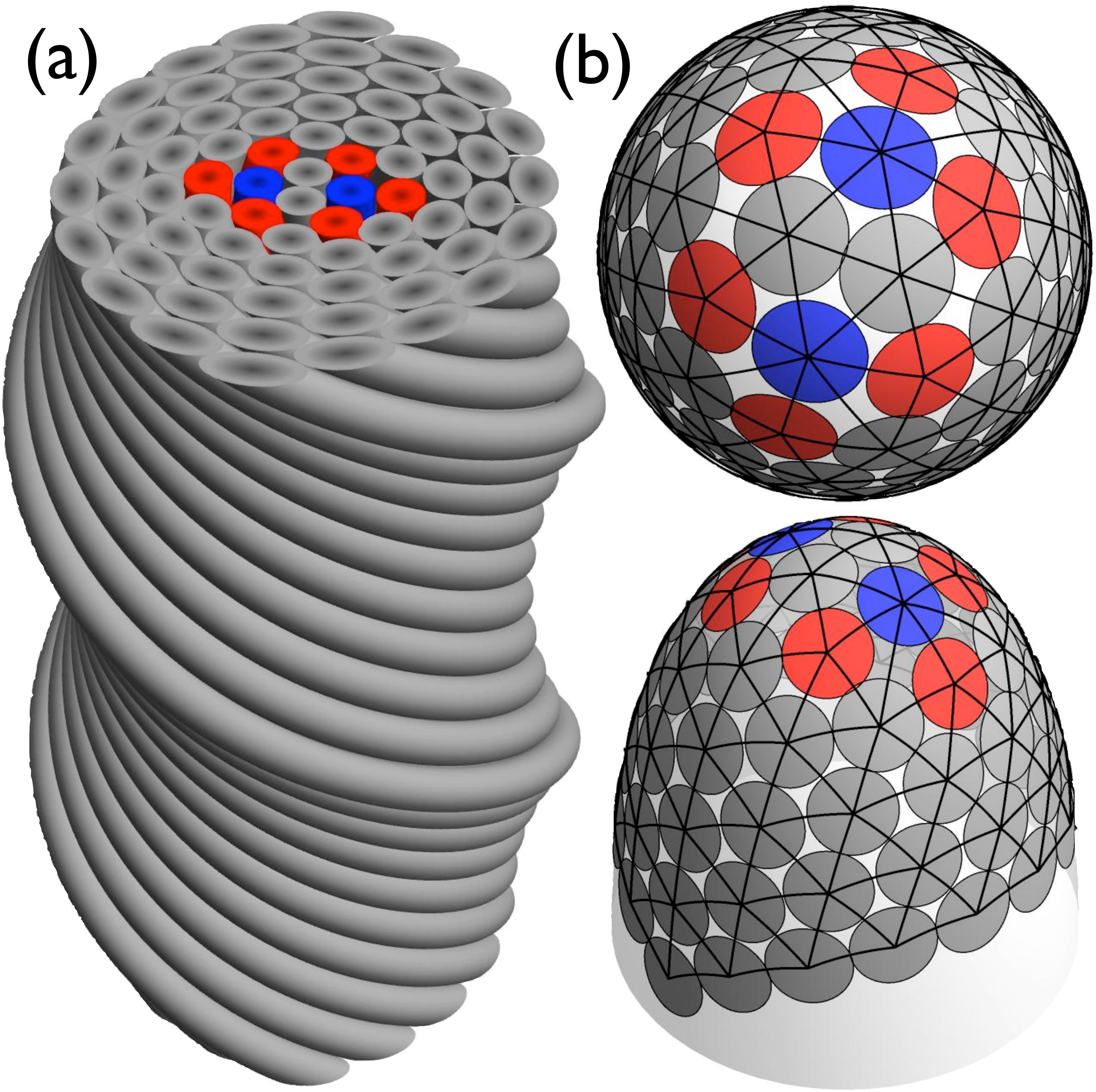}
\caption{(a) A bundle with 70 filaments and $\theta = 57.7^{\circ}$. (b) Top and side views of the corresponding disc packing on the bundle-equivalent surface, which shares the packing geometry and topology with the twisted bundle. 5 and 7-fold coordinated elements are colored red and blue respectively.}
\label{fig:BundleDome}
\end{figure}

The universal twist dependence of $Q$ in ground-state bundles derives from a formal mapping of the inter-filament distance and the geodesic distances between points on a dome-like surface that encodes the metric properties of bundles. The details of this mapping can be found in Appendix \ref{sec:DomeDerivation}. To summarize, the positions of filament centers in the cross section of a twisted bundle can be mapped exactly to positions on what we will refer to as the \textit{bundle-equivalent surface}. This surface is spherical at the top---corresponding to the small $\theta$ center of the bundle---and tapers to a cylindrical geometry along its height---corresponding to the large $\theta$ exterior of the bundle. Starting at the center of a bundle and traveling along an outward radial path, is equivalent to starting at the top of the dome and traveling outward along the surface. An example of a bundle packing mapped in this way is shown in Fig.\ \ref{fig:BundleDome}.

The mapping between the interfilament distance in twisted bundles and the geodesic distance on the dome-like surface, implies that the packing of the bundles can be understood in terms of the packing of point-like objects on the a curved surface, for which classical theorems of differential geometry relate the geometry of a surface to the topology of the nearest-neighbor bond network \cite{Bowick2009}. A finite-diameter bundle corresponds to a finite ``crystalline" patch of particles or discs extending over a portion of the dome-like surface, centered on the high-curvature central pole. In Appendix \ref{sec:DomeDerivation}, we derive a simple generalization of the Euler-Poincar\'e theorem for a triangular bond-network on a curved 2D surface that relates the topological charge $Q$ of the packing to the integrated Gaussian curvature over the surface domain covered by the patch. Based on the simple assumption that large deviations from equilateral packing at the boundary are energetically expensive, and therefore, unlikely in ground-state packings, we exploit this theorem to derive an expression for $Q_{id}$, the {\it ideal} value of topological charge of the interior packing, which requires no distortion from equilateral bond-order at the patch edge (and equivalently, the bundle surface). Assuming an axisymmetric shape for the boundary of the packing, evaluating the integrated curvature on the dual surface gives the ideal charge purely in terms of twist angle,
\begin{equation}
Q_{id} = 6 \big( 1 - \cos^3 \theta \big) .
\label{eq:QidSim}
\end{equation}
Importantly, the $\theta$-dependence of $Q_{id}$ encodes the increase of integrated Gaussian curvature as the patch size grows large compared to the curvature radius of the dual surface (proportional to the pitch). As the surface domain grows to cover a larger portion of the curved surface, the preferred topological charge becomes non-zero. Though the presence of the free boundary of the bundle allows the topological charge to adjust based on purely energetic considerations, the positive curvature of the bundle-equivalent surface suggests a connection between the optimal packings of highly twisted bundles and the better-known packings of particles (or discs) on closed, spherical surfaces, studied in the context of the generalized Thomson problem \cite{Bowick2002, aste2008pursuit, Altschuler1997}. In the language of the continuum theory of curved 2D crystals, excess 5-fold disclinations {\it screen} the elastic stresses generated by geometric frustration, such that increasing curvature (or twist in the case of bundles) requires an increasing number of ``neutralizing" disclinations.

\begin{figure}[h]
\centering
\includegraphics[width=8.0cm]{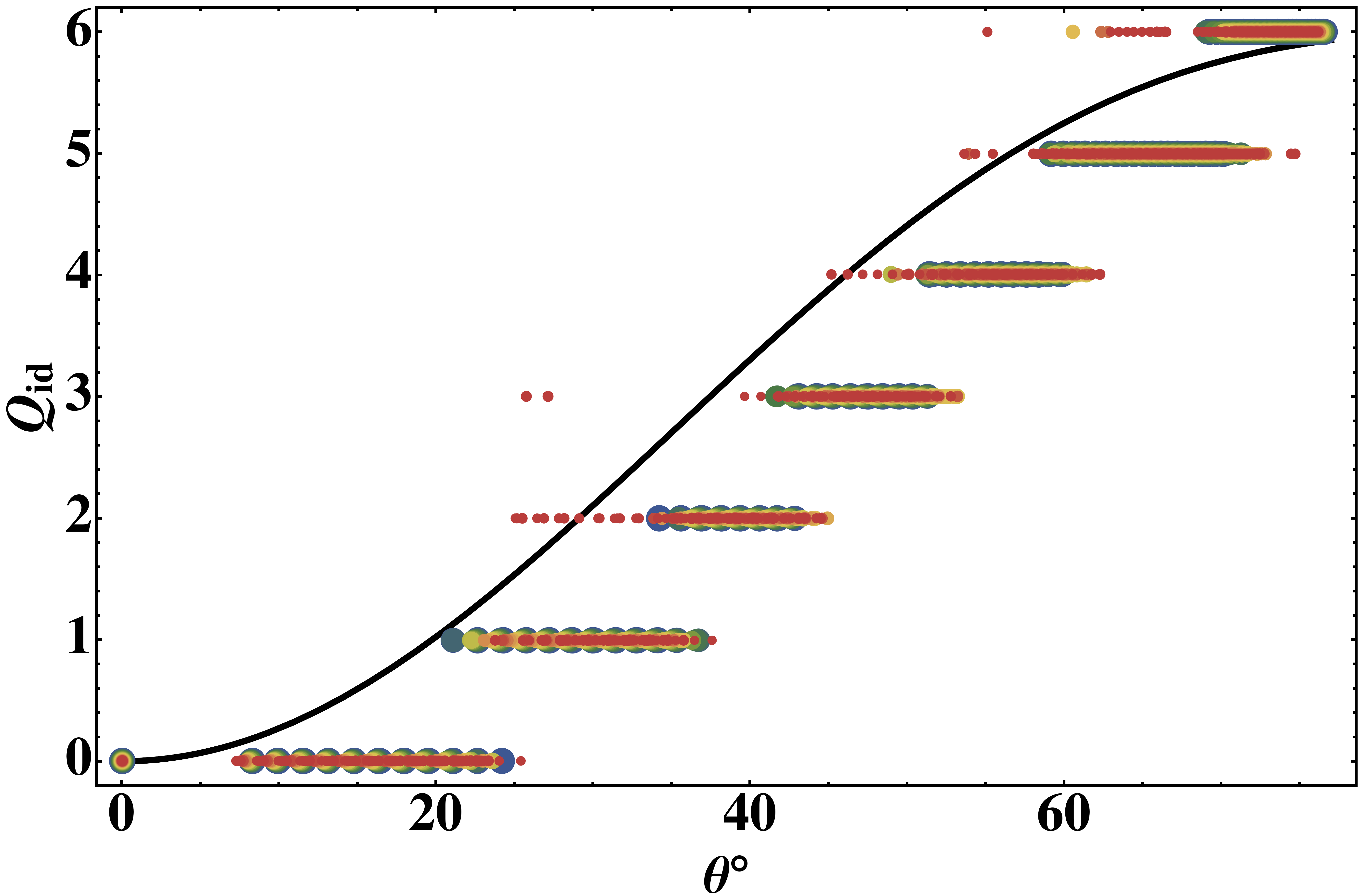}
\caption{The ideal charge, $Q_{id}$, from eqn (\ref{eq:QidSim}) (solid black line) vs twist angle $\theta$. The data points are the $Q$ values for an accumulation of ~5000 simulation results for the various twist angles and bundle sizes introduced in section \ref{sec:NumSims}. The number of filaments range from $N=16$ (red data points) to $N=196$ (blue data points).}
\label{fig:QidSim}
\end{figure}

Though $Q_{id}$ increases continuously from 0 at $\theta=0^{\circ}$ to 6 at $\theta = 90^{\circ}$, the actual topological charge of the packing may only take on integer values, the simple assumption that the integer value of $Q$ in ground-state packings is determined by the {\it closest} integer value to $Q_{id} (\theta)$ is remarkably consistent with our numerical simulations of twisted bundles. In Fig.\ \ref{fig:QidSim} we plot the $Q$ for bundle simulations vs. twist angle and compare this to the continuously increasing value of $Q_{id}(\theta)$.
While the agreement is between $Q$ and $Q_{id}$ is imperfect, this above argument highlights the fundamentally geometric nature of packing frustration in twisted bundles, and importantly, provides a natural explanation for the observation that optimal values of $Q$ are independent of $N$, determined only by $\theta$, which controls the value of the integrated curvature on the bundle-equivalent surface.

This mapping of filament packing in bundles to a spherically-curved surface not only explains the increase of $Q$ with twist, but it also sheds light on a basic tendency of increasing the number of disclinations per topological charge with bundle size, as observed in our simulated ground states. In structures where $N_{Disclination}  = Q+ 2n$, a defect ``complex" can be described as a charged {\it scar}, usually observed as chain of alternating 5 and 7-fold disclinations. Importantly, the transition from compact (point-like) to extended (string-like) disclination defects has been predicted and studied in great detail in the context of crystalline order on spheres~\cite{Bowick2000,Bowick2002,Bowick2009}. Adapting the scaling arguments for stability of dislocations (neutral 5-7 pairs) in the vicinity of a scar on a sphere~\cite{Bowick2000}, we expect the number of 5-7 pairs decorating a scar in a bundle to grow as $\sim R_0/d \propto \sqrt{N}$ at fixed value of twist, a trend that is in reasonable agreement with our simulations.

\subsection{Twist-dependence of cohesive energy: Infinite-length limit}
\label{sec:DiscreteEnergy}
\begin{figure}[h]
\centering
\includegraphics[width=8.0cm]{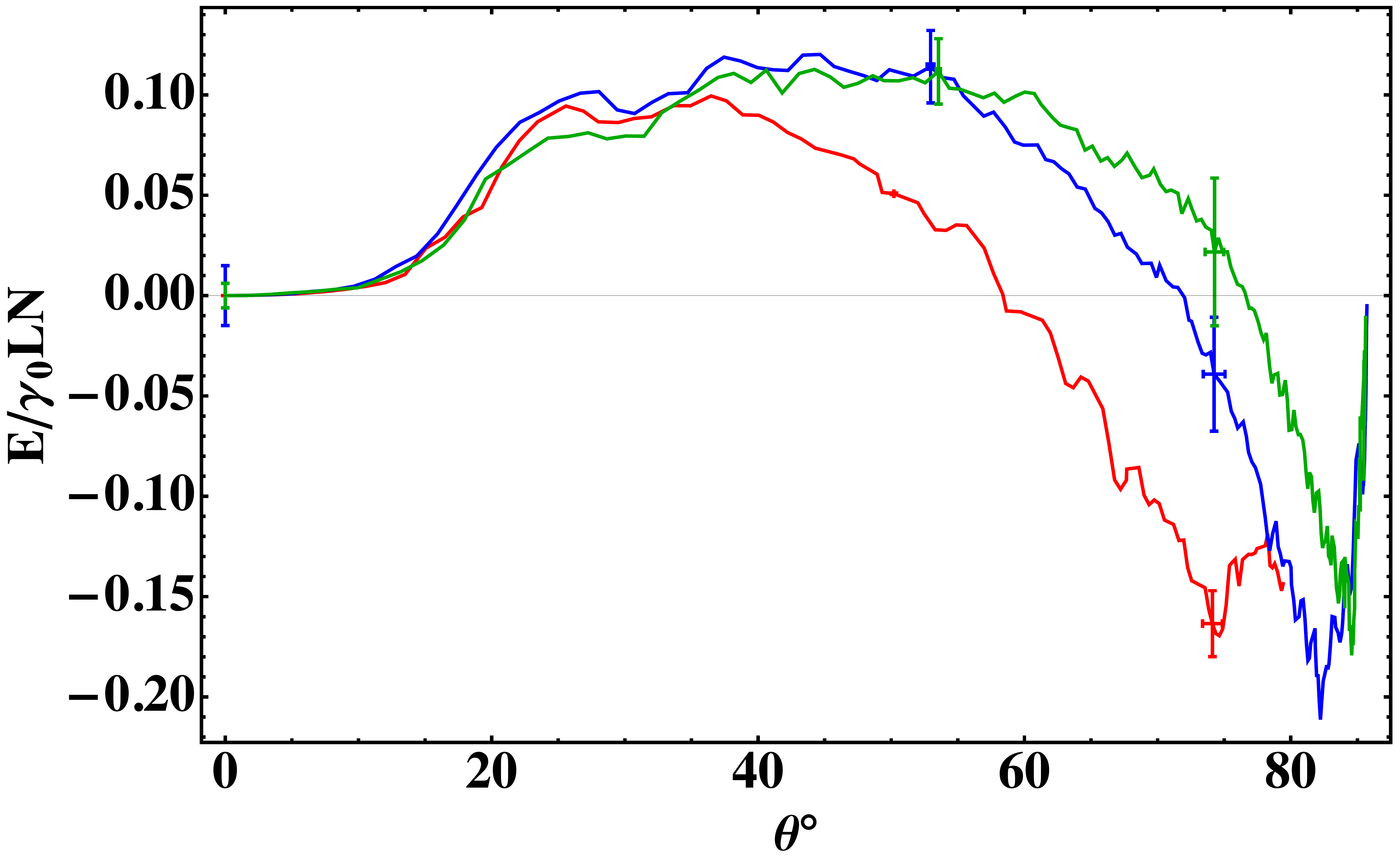}
\caption{Mean filament cohesive energy per unit length vs twist angle, $\theta$, for three selected bundle sizes: small $N = 34$ (red), medium $N = 82$ (blue), and large $N = 184$ (green). The roughness is a consequence of sudden rearrangements in the packings to accommodate the twist-induced geometric frustration. In this and all following plots, the energy change is defined relative to the {\it untwisted} case. Error bars are shown for each $N$ at three select values of $\theta$. These error estimates derive from the standard deviation taken from 100 implementations of the numerical ground-state search algorithm (each of which samples $\sim10^3$ initial configurations). These estimates suggest our sampling yields bundle energies to within less than 1\% of the true ground-state energy.}
\label{fig:TotalSimEn}
\end{figure}

Having analyzed the twist-dependence of the structure of minimal-energy bundles in terms of the topological charge of the packing, we now consider the twist-dependence of cohesive energy, with the aim of discerning the influence of the universal evolution of $Q$ on energy. We find that in the large-$N$ limit, bundle energetics converge to a common behavior. Here, we focus on thermodynamic behavior of small, intermediate and large bundles in detail. Fig.\ \ref{fig:TotalSimEn} shows the change in mean cohesive energy per filament length for these three cases.
We observe a common behavior for each of the bundle sizes. Twisting initially increases the energy of untwisted bundles, until reaching a rough plateau region at intermediate twist. However, further twist {\it lowers} the cohesive energy, ultimately driving it {\it below the energy of the untwisted state}. Notice that the crossover angle shifts to larger $\theta$ with increasing $N$, an effect that we ultimately attribute to the decreasing fraction of filaments at the surface of larger bundles (see section \ref{sec:SurfEnergy}).

\begin{figure}[h]
\centering
\includegraphics[width=8.0cm]{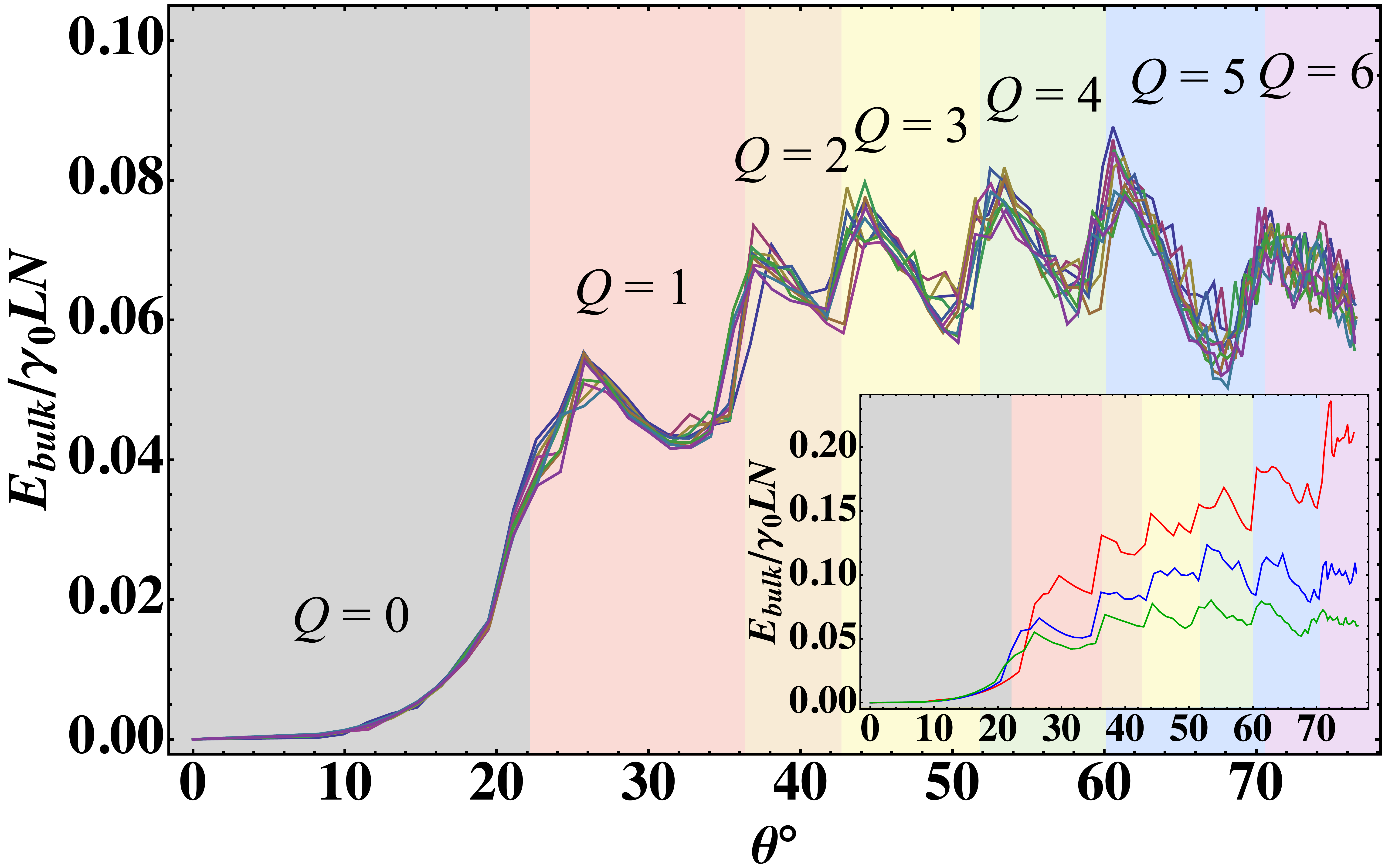}
\caption{Bulk cohesive energy vs. twist angle for ten values of $N$ from 166 to 193, showing convergence a of internal energy for large $N$. The discrete series of minima correspond to discrete increases in $Q$. Background colors correspond to the trends in $Q$ presented in Fig.\ \ref{fig:SimPhaseD}. Inset shows three bundle sizes: small $N = 34$ (red), medium $N = 82$ (blue), and large $N = 184$ (green).}
\label{fig:InternalSimEn}
\end{figure}

In the next section, we carefully analyze the surface geometry of twisted bundles to show that the tendency to decrease the cohesive energy with twist is driven by a decrease of non-contacting filament lengths at the boundary of long bundles. In this section, we show that removing the effects of changes in filaments at the boundary of the bundle reveals a universal dependence of the bulk packing on twist. To perform this analysis, we subtract the surface energy of the bundle, which is defined formally in section \ref{sec:ContinuumSurfEn}, from the total energy of the discrete model to define the {\it bulk cohesion energy}, $E_{bulk}$. Fig.\ \ref{fig:InternalSimEn} shows the change in mean $E_{bulk}$ per filament length for large bundles in the range of $N=166-193$, revealing a common increase of energy relative to the untwisted state.
For small twist, the bulk energy increases smoothly with $\theta$, due to the increasing frustration of inter-filament spacing in defect free bundles. The small-$\theta$ dependence of large-$N$ bundles is consistent with results of elasticity theory calculations which show that $E_{bulk} \sim \theta^4$ in defect-free bundles \cite{Grason2012, Grason2010}. Also consistent with elasticity theory results, is the appearance of cusps in $E_{bulk}$ at transitions in topological charge, such as the transition from $Q=0$ to $Q=+1$ at $\theta \simeq 27^\circ$. Just beyond the transition, energy decreases with twist, highlighting the key ability of excess disclinations to mitigate the twist-induced frustration, taming the rapid growth in bundle energy. The five local minima mark values of twist that are locally stable due the optimal screening provided by discrete values of $Q=1,2,3,4$ and 5.

We consider the convergence to the large-$N$ bulk energy in the inset of Fig.\ \ref{fig:InternalSimEn}, which shows the twist dependence of $E_{bulk}$ (per filament) for the small, intermediate and large bundles. Significantly, at a given value of twist, the bulk per filament cost of twist decreases as $N$ increases. Underlying the $N$-dependence for small bundles are two distinct types of energies counted in $E_{bulk}$. The first contribution is the far-field elastic cost of geometric frustration and excess disclinations in twisted bundles. Based on the continuum theory analysis of refs.\ \cite{Grason2012, Grason2010, Azadi2012}, these $\theta$-dependent costs are proportional $N$ and determined only by $Q$, consistent with the universal large-$N$ behavior of $E_{bulk}$. The second type of contribution may be attributed to the ``core energy" of the excess disclinations appearing in the packing above the critical twist. Core energies represent corrections to the continuum theory behavior due to, for example, the fewer cohesive contacts at the core of 5-fold disclinations, as well as the large-strain deviations in the neighborhood of defects. These costs grow in proportion to the total number of defects in the bundle, and hence amount to a larger proportion of $E_{bulk}$ for smaller bundles. As the total number of defects does not grow faster than $N$, the defect cores contribute a negligible amount to the energy density $E_{bulk}/N$ relative to the far-field elastic costs of disclinations and twist as $N \to \infty$.

To summarize, this analysis of the bulk cohesive energy of ground-state packings of our discrete filament model reveals two key influences of twist on the cohesive energy of bundles. First, twist frustrates the uniform inter-filament packing allowed in straight bundles, leading to a necessary increase in energy with twist angle. Second, we find that increasing twist triggers the stability of excess 5-fold disclinations in the cross-sectional packing that mitigate the growth of bundle energy associated with twist-induced frustration of nearest-neighbor spacing. In the next section, we consider the twist-dependent changes in cohesive filament contact at the surface of bundles, which for sufficiently long bundles evidently counteract the increases in $E_{bulk}$ with twist.

\section{Surface energy of twisted bundles}
\label{sec:SurfEnergy}
The ground-state results of our discrete model simulations show that while the geometric frustration leads to a cohesive energy costs in the bulk of twisted bundles, there is a competing tendency for twist to lower this energy, eventually even to values lower than the energy of the untwisted state (seen in Fig.\ \ref{fig:TotalSimEn}). In this section, we show that this driving force for twist results from an overall decrease in the surface energy of the bundle. The surface energy accounts for the deficit of cohesive contact at the periphery of bundle, which we analyze in the continuum limit of vanishing filament diameter to show that twisting decreases the total number and length of exposed filaments at the sides and ends of the bundle. We then compare this to the results of our discrete model simulations. Finally, we pair the surface driving force for twist in our continuum model to the mechanical cost for bending filaments, to determine the thermodynamic preferences for twist in cohesive bundles in terms of the bundle aspect ratio and filament stiffness.

\subsection{Surface geometry of twisted bundles}
\label{sec:SurfGeometry}
In this section, we analyze the surface geometry of twisted filament bundles in the continuum limit where filament diameter is small compared to both filament length and the lateral size of the bundle, with the goal of developing analytical formula for the dependence of surface energy on twist. In this spirit, we make the additional approximations that: 1) the filament packing is locally hexagonally-close packed, with a density that is independent of twist; and 2) the shape of the bundle is axisymmetric, with an outer cylindrical radius, $R$. The first approximation is clearly violated in the neighborhood of disclinations that enter the packing at finite twist. Though, for large bundles $N\gg1$, the local packing is non-hexagonal for only a relative minority of filaments. Hence, we assume that the local filament spacing and occupied volume fraction change only modestly in the bulk of twisted bundles, which is consistent with the density of maximally-compact bundles studied in ref.\ \cite{Bruss2012}. For straight filament bundles, the second approximation (cylindrical bundle symmetry) clearly fails to capture the hexagonal faceting of the bundle sides. However, our simulations show that bundles become more axisymmetric at high twist as the packing trades high-energy corners at the bundle surface for excess disclinations in the bundle interior~\footnote{This follows from the fact that the sum of topological charge interior to the bundle and the net deficit of nearest neighbor contacts at the boundary are constrained to obey, $Q+\sum_n (4-n)V_b(n)=6$. See, e.g.\ \cite{Bowick2009}.}. 

\begin{figure}[h]
\centering
\includegraphics[width=8.0cm]{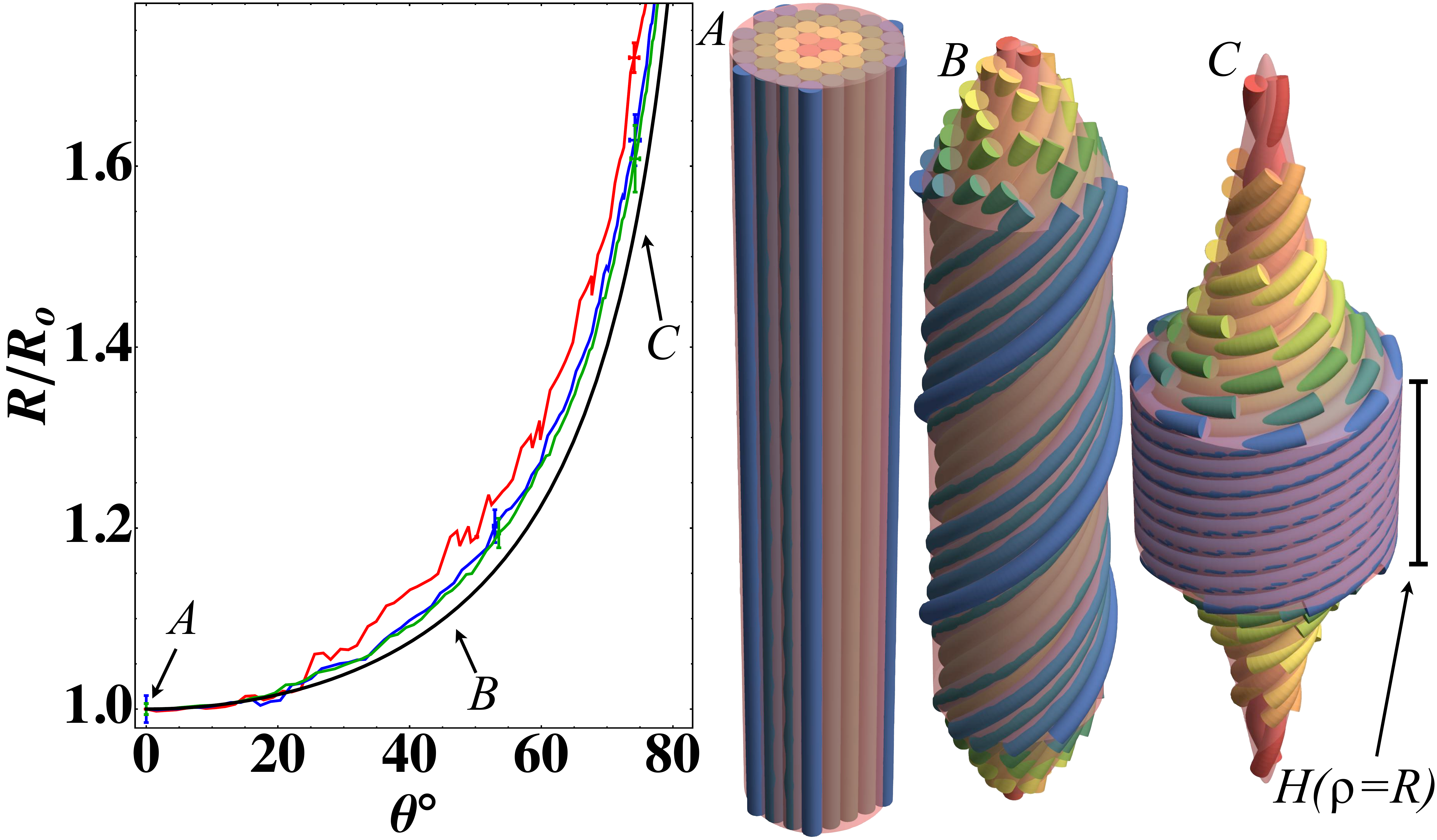}
\caption{Filament bundle radius, $R/R_{0}$ vs twist angle, $\theta$ from numerically simulated ground states. The black line is the continuum model prediction of eqn (\ref{Rcont}), while the discrete model is represented with three bundle sizes: small $N = 34$ (red), medium $N = 82$ (blue), and large $N = 184$ (green). Both the continuum approximation (opaque pink surface, radius is $R + d$ to account for the filament diameter), and the discrete model representations are shown together for three example twist angles for a bundle of $N = 46$. The height of the outermost filaments is shown for example $C$. Error bars are shown for each $N$ at three select values of $\theta$, these represent the standard deviation (for 100 trials) of the low energy state found via the ground-state search algorithm.}
\label{fig:RwTwist}
\end{figure}

Based on these assumptions, we now consider the change in the surface shape with twist (shown in Fig.\ \ref{fig:RwTwist}). 
As twist increases, the helical tilt of filaments away from the center of rotation increases as $\theta(\rho) =\arctan( \Omega \rho)$. Since the contour length of filaments are fixed to $L$, twist requires a change of the height, $H$, the extent of a filament along the pitch axis, according to
\begin{equation}
H(\rho) = L \cos \theta (\rho) = \frac{ L}{\sqrt{1+(\Omega \rho)^2} } .
\label{eq:FilamentZHeight}
\end{equation}
Hence, for non-zero $\Omega$, the vertical profile of the ends varies with radius. Assuming filaments distribute the taper equally over both of the free ends of the bundle, the shape of this tapered profile is described by $H(\rho)/2$, as seen in the example bundles in Fig.\ \ref{fig:RwTwist}. It is important to point out that the curved shape of this profile is unrelated to the geometry of the bundle-equivalent surface which encodes the metric properties of interfilament spacing (see section \ref{sec:Mapping}).

Local decreases in the height of the bundle with twist imply that the lateral radius must necessarily increase in order to preserve a constant volume and density. Given $\Omega$ and an outer bundle radius, $R$, the volume within a bundle is easily computed as $V(\Omega, R) = 2 \pi \Omega^{-2} L \big(\sqrt{1+ (\Omega R)^2} -1 \big)$. Assuming an untwisted bundle radius of $R_{0}$, the $\Omega$-dependence of $R$ is determined from the solution to $V(\Omega, R) = \pi R_{0}^2 L$,
\begin{equation}
\label{Rcont}
R = R_{0} \sqrt{1 + (\Omega R_{0}/2)^{2}}.
\end{equation}
This formula, while derived from global considerations of volume conservation, implicitly encodes the same constraints of lateral filament-packing in twisted bundles described by the mapping to the dome-like surface: twisting a bundle reduces the number of filaments that can be packed at a given radius $\rho$ by $\cos \theta(\rho)$, implying that filaments must be redistributed to larger radii, and notably compares quantitatively with simulated bundle radii, as shown in Fig.\ \ref{fig:RwTwist}.

\subsection{Continuum-limit surface energy}
\label{sec:ContinuumSurfEn}
We now proceed to analyze the contribution to the cohesive energy deriving from the twist-induced changes of filament contact at the surface of the bundle. The surface energy per unit area, $\Sigma$, accounts for the loss of favorable cohesive interactions due to exposure of non-contacting filament length at the surface. Simply put, the surface energy attributes a loss of cohesive energy, $\gamma_{0}/2$, per unit length of lost neighbor contact, relative to the hexagonal packing of the bulk. Following standard arguments for surface energy, the factor of $1/2$ follows from the fact that the separation of one contacting pair creates two non-contacting filaments \cite{maugis2000contact}.

\begin{figure}[h]
\centering
\includegraphics[width=8.0cm]{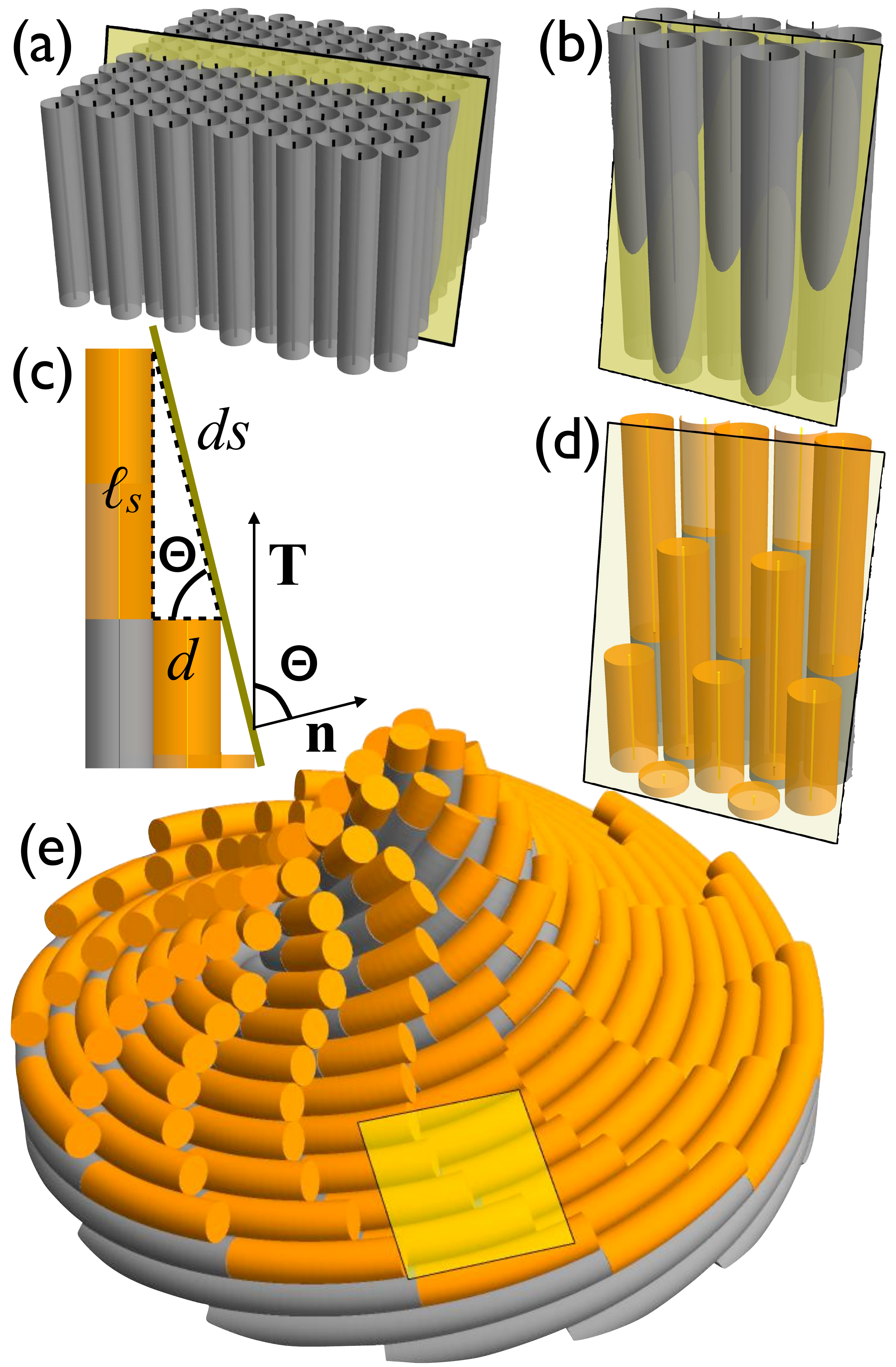}
\caption{(a) A bulk collection of filaments with a cutting surface. (b) Zoomed in view. Oblique (c), and side (d) views of the surface cut. The orange segments represent the lengths of filaments that are now no longer interacting with the neighbors in front of them. This length, $\ell_{s}$, is dependent on the angle, $\Theta$, between the surface normal, ${\bf n}$, and the filament tangent vector, ${\bf T}$. This cut corresponds to an end surface area section, such as the example shown in (e).}
\label{fig:BulkCuts}
\end{figure}

The surface of a twisted bundle is composed of exposed filaments that lack the full complement of neighbors to achieve the maximal cohesive energy density. In a twisted filament bundle, non-contacting filament lengths arise in two ways. First, filaments at the radial sides expose lengths of non-contact along the entire outer contour. Second, due to the finite contour length of filaments, twist leads to ``slip" of filament pairs at the ends of bundles. The cohesive energy cost in both cases may be derived by considering the creation of non-contacting filament length by introducing a planar ``cut" through a bulk hexagonal array of filaments, as shown in Fig.\ \ref{fig:BulkCuts}a. The per area exposure of non-contacting filament length is determined by the unit normal ${\bf n}$ to the planar cut (the normal of the free surface element) and the local orientation of filament tangents, ${\bf T}$. Consider, for example, the loss of contact, i.e.\ the slip length, $\ell_s$, for the neighbor pair shown in Fig.\ \ref{fig:BulkCuts}c, where the tilt of the cut direction is along the neighbor separation. In this case, it is straightforward to relate the length of surface separating filament ends, $ds$, to slip length, $\ell_s = ds |\sin \Theta|$, where $|\sin \Theta| = |{\bf T} \times {\bf n}|$.
In general, summing over the slip of nearest-neighbor contacts yields a surface energy per unit area of the form,
\begin{equation}
\Sigma = \alpha \frac{\Sigma_{0} }{2 }|{\bf T} \times {\bf n}|,
\label{eq:SurfEnergy}
\end{equation}
where $\Sigma_{0} = \gamma_{0}/d$, and $\alpha$ is a numerical coefficient deriving from the orientation of neighbor directions with respect to the surface element. For example, for the low-energy sides of the bundle where filaments are perpendicular to the exposed surface ($|{\bf T} \times {\bf n}| = 1$), it is straightforward to show that $\alpha = 2$, due to the two fewer neighbors for filaments at the surface relative to the bulk. A more detailed calculation (given in Appendix \ref{sec:OrentationAlpha}) that averages the slip-cost of a cut hexagonal array with respect to all cutting directions yields $\alpha = 4 \sqrt{3}/\pi \simeq 2.2$. For the remainder of the article, we neglect the variation in relative geometry of neighbor directions and cut directions and simply take $\alpha = 2$ for all bundle surfaces.

For long filaments, the radial sides of the bundle carry most of the surface energy as filaments are normal to the free surface along their lengths and $|{\bf T} \times {\bf n}|=1$ is maximal. We define $E_{side}$ as the change in surface energy relative to the untwisted bundle, which we calculate using eqns (\ref{eq:FilamentZHeight}) and (\ref{Rcont}),
\begin{equation}
E_{side} = \Sigma_{0} A_{0}\left( \frac{ R/R_{0}}{\sqrt{1+(\Omega R)^{2}}} - 1\right), 
\label{eq:SideEn}
\end{equation}
where $A_{0} = 2 \pi R_{0} L$ is the side area of the untwisted bundle. Since the height of bundle side, $H(R)$, decreases more rapidly with twist than the lateral growth in radius, $E_{side}$ is a decreasing function of twist. For small twist, $E_{side} (\Omega R \ll 1) \simeq -(3/8) \Sigma_{0} A_{0} (\Omega R)^2$, indicating that any finite measure of twist reduces surface exposure filaments at the sides. Since the length of non-contacting filaments at the boundary is fixed to $L$, this change must derive from a change in the number of surface filaments. This demonstrates that a twisted bundle contains a larger proportion of its filaments in the interior than does an untwisted bundle.

\begin{figure}[h]
\centering
\includegraphics[width=8.0cm]{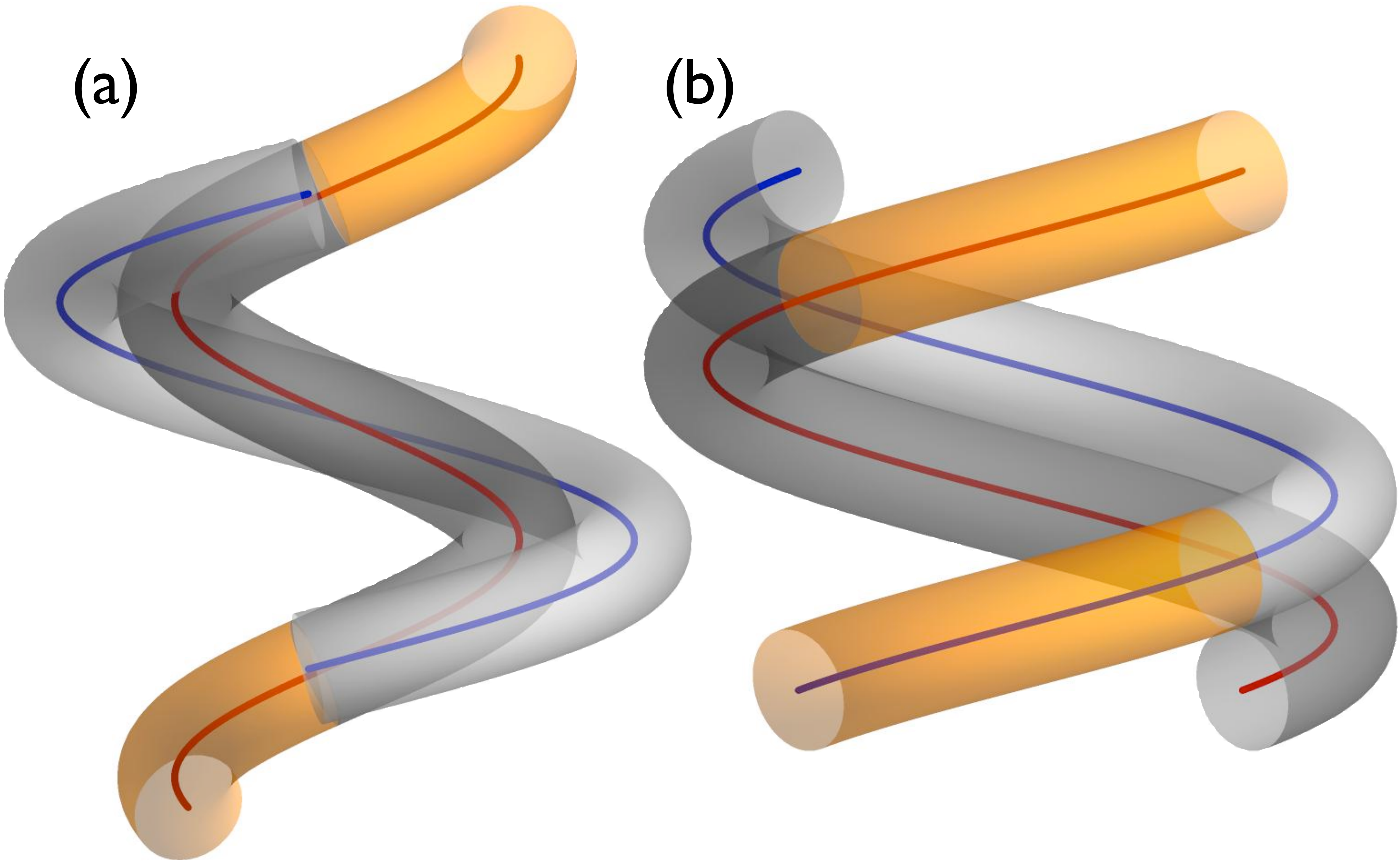}
\caption{An interacting pair of filaments with two possible orientation. (a) {\it Radial slip} - both filaments share the same azimuthal position, $\phi$. Here, the inner filament loses contact at both ends. (b) {\it Azimuthal slip} - both filaments share the same radius, $\rho$. Here, each filament experiences reduced contact length at one end of the bundle. }
\label{fig:SlipLenDia}
\end{figure}

While twist reduces $E_{side}$, this reduction comes at the expense of increasing surface exposure of non-contacting length at the {\it ends} of the bundle. We evaluate the surface energy contribution from one of the ends of the bundle, $E_{end}$, beginning with eqn (\ref{eq:SurfEnergy}). The normal and tangent vectors are defined as
\begin{align}
{\bf n} = \frac{\hat{z} + H' /2 \hat{\rho}} {\sqrt{1+(H'/2)^2}}  \\
{\bf T} (\rho) = \cos \theta(\rho) \hat{z} + \sin \theta(\rho) \hat{\phi}.
\end{align}
Integration over ends of the surface area of ends, for which $dA = 2 \pi d \rho~ \rho  \sqrt{1+(H'/2)^2}$ (where $H' = \partial_\rho H$), yields
\begin{align}
E_{end} &= \Sigma_{0} \int_{end} dA |{\bf T} \times {\bf n}| \nonumber \\
&= 2 \pi \Sigma_{0} \int_{0}^{R}\frac{ d\rho | \Omega| \rho^{2} }{\sqrt{1 + (\Omega \rho )^{2} }} \left( 1+ \frac{ (\Omega L/2)^2}{ \left[1 + (\Omega \rho )^{2} \right]^2 } \right)^{1/2}.
\label{eq:EndSurfEn}
\end{align}
Analysis of the integrand of eqn (\ref{eq:EndSurfEn}) reveals that $E_{end}$ has two analytically tractable limits whose form depend on the relative magnitude of $\Omega L$, which is proportional to the number of helical turns of a bundle, and $\sec^2 \theta= 1+(\Omega R)^2 $. In the limit of infinite length (and finite twist) the surface energy per end becomes,
\begin{multline}
\label{Endlong}
\lim_{\Omega L \gg \sec^2 \theta } E_{end} =  \pi L \Omega^{-1} \Sigma_{0}\big[{\rm arcsinh}(\tan \theta) - \sin \theta \big],
\end{multline}
which has a small-$\Omega R$ behavior $E_{end} (\Omega L \gg \sec^2 \theta) \simeq \frac{ A_{0} \Sigma}{6} (\Omega R)^2$~\footnote{Strictly speaking, this is the limit $L \gg \Omega^{-1} \gg R_{0}$, since $\Omega R \gg1$.}. In this limit, the surface energy of ends derives predominantly from {\it radial slip} of neighbor filament pairs extending to different heights, for small twist, $\ell_s({\rm rad}) \sim d |H' |\sim  d \Omega^2 R L$, per pair (see Fig.\ \ref{fig:SlipLenDia}a).
In the opposite, limit of vanishing length the surface energy takes the form,
\begin{equation}
\label{Endshort}
\lim_{\Omega L \ll \sec^2 \theta} E_{end} = \pi R \Omega^{-1} \Sigma_{0} \Big[ \cot \theta - \frac{ {\rm arcsinh}( \tan \theta) }{ \tan \theta} \Big],
\end{equation}
which has a small-$\Omega R$ limit, $E_{end} (\Omega L \ll 1) \simeq \frac{ \pi R^2}{3} |\Omega R|$. The end surface energy cost in this limit (few helical turns per bundle) is dominated by interfilament slip between {\it azimuthally-separated} neighbors, for which $\ell_s ({\rm azi} )\approx |{\bf T \cdot \hat{\phi}} | d \sim d |\Omega R|$ at small twist (see Fig.\ \ref{fig:SlipLenDia}b).

\begin{figure}[h]
\centering
\includegraphics[width=8.0cm]{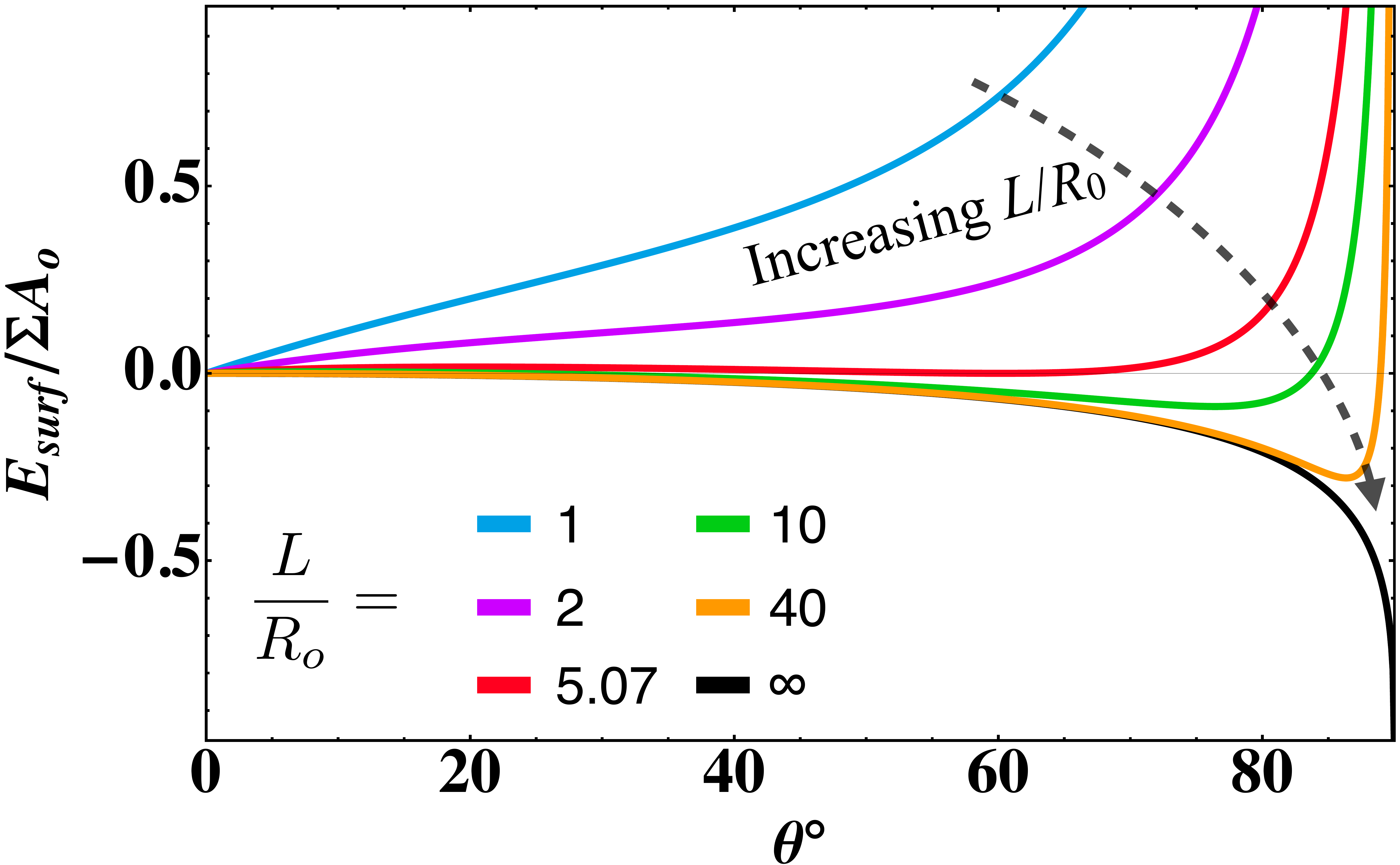}
\caption{Filament bundle surface energy vs twist angle for various aspect ratios of $L/R_{0}$.}
\label{fig:SurfaceEnergy}
\end{figure}

Notably, the ratio of the surface energy contributions captured in eqns (\ref{Endlong}) and (\ref{Endshort}), which derive from the two distinct modes of slip occurring at the ends of twisted bundles, are consistent with the relative magnitudes of {\it radial} vs. {\it azimuthal} slip in weakly twisted bundles, $\ell_s({\rm rad}) / \ell_s ({\rm azi} ) \sim |\theta | (L/R_{0})$. This implies that the {\it aspect ratio} of the bundle, $L/R_{0}$, is a key parameter governing the twist-dependence of surface energy. In Fig.\ \ref{fig:SurfaceEnergy} we plot the total surface energy, $E_{surf}=E_{side} + 2 E_{end}$, as a function of twist angle $\theta$ for aspect ratios ranging from $L/R_{0} \to 0$ to the infinite length limit, $L/R_{0} \to \infty$. 

First, we consider the behavior of the $L/R_{0} \to \infty$ limit, where the decrease with twist in surface contact at the sides of the bundle overwhelms the additional cost of radial slip described by eqn (\ref{Endlong}), such that the surface energy is {\it unstable} to twist and obtains its minimum for the maximally twisted state $\theta \to 90^{\circ}$. In the opposite limit of $L/R_{0} \ll 1$ (vanishing aspect ratio), the side and radial-slip contributions at the ends contribute negligibly to $E_{surf}$ since both scale with contour length of filaments. Thus the surface energy is dominated by {\it positive} cost of exposing {\it azimuthal slip} at the ends of the bundle, eqn (\ref{Endshort}). For small twist, this cost increases linearly as $E_{end} (L/R_{0} \ll 1)/ A_{0} \sim|\theta| (R_{0}/L) $ and diverges for large twist, $E_{end} (L/R_{0} \ll 1)/ A_{0} \sim \cot^{1/2} \theta (R_{0}/L)$ such that the minimal surface energy corresponds to the untwisted state. For intermediate aspect ratios we see a transition between these extremes, with the minimal surface energy occurring at $\theta = 0^{\circ}$ for short bundles, and jumping to finite and large values of twist above a critical aspect ratio $L/R_{0} \simeq 5.07$.

\begin{figure}[h]
\centering
\includegraphics[width=8.0cm]{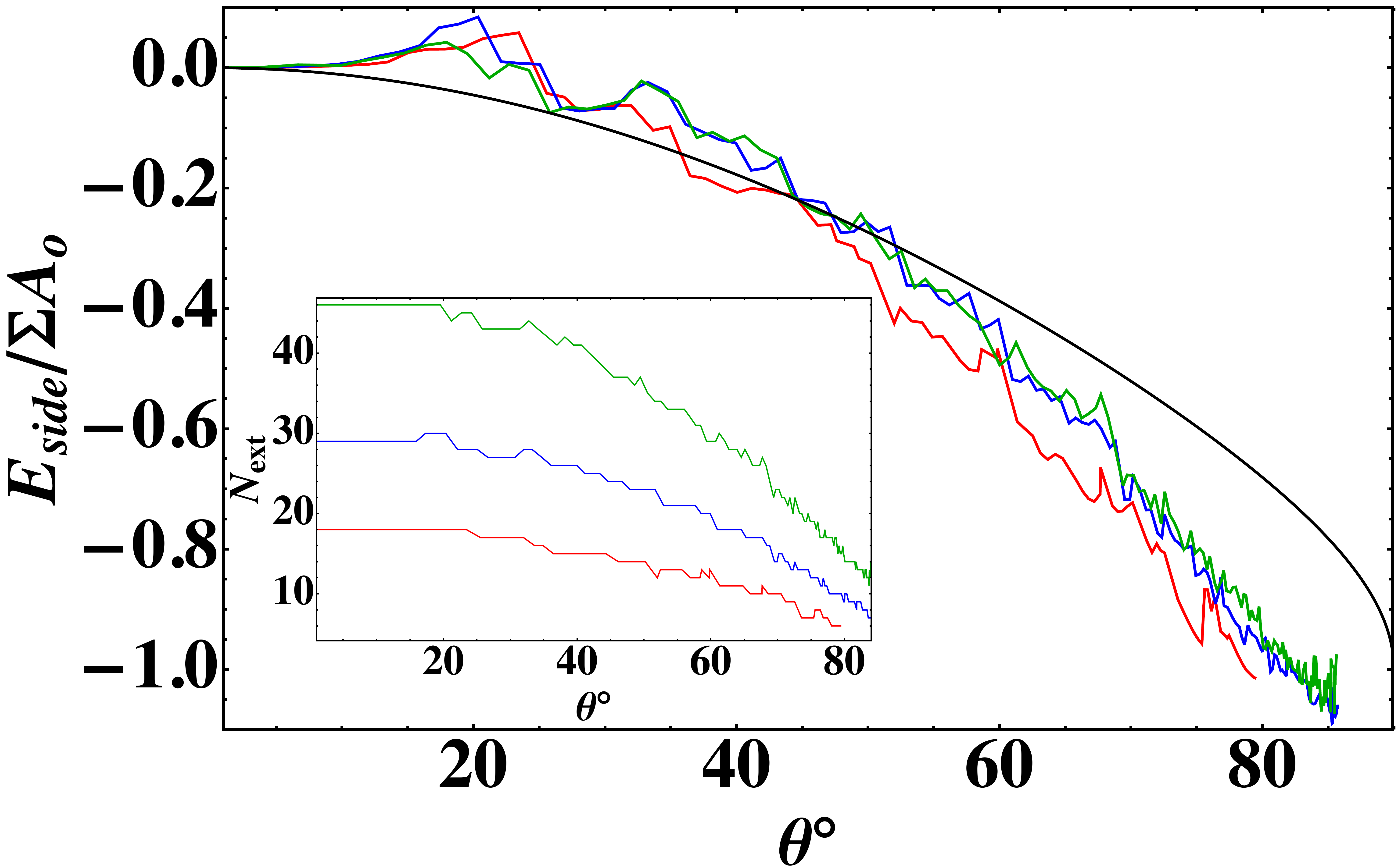}
\caption{External filament cohesive energy vs twist angle for three selected bundle sizes: small $N = 34$ (red), medium $N = 82$ (blue), and large $N = 184$ (green). Black line is the continuum model expression eqn (\ref{eq:SideEn}). Inset shows number of external filaments, $N_{ext}$ vs twist angle.}
\label{fig:ExternalSimEn}
\end{figure}

We assess the quality of the continuum-limit surface energy analysis by direct comparison to the numerical simulations of the discrete filament model. Simulations of bundle cross sections in section \ref{sec:NumSims} are carried out in the $L/R_{0} \to \infty$ limit so that surface energy changes with twist are derive only from $E_{side}$ and the radial-slip contributions to $E_{ends}$, which are both proportional to $L$. To extract the surface energy of bundle sides, we calculate the excess energy of surface filaments due to fewer filament neighbors than the predominantly six-fold packing in the bulk,
\begin{equation}
\label{Es/dis}
E_{side} =  \frac{L}{2} \sum_{i\in b}\Big[ \sum_{j \neq i} \gamma(\Delta_{ij}) - 6 \gamma(d) \Big] ,
\end{equation}
where $i \in b$ refers to filaments at the surface of the bundle. Fig.\ \ref{fig:ExternalSimEn} compares the relative change in surface energy at the sides of small, intermediate and large bundles in our discrete model, to the continuum expression for $E_{side}$ is eqn (\ref{Es/dis})~\footnote{For the discrete model, the value of $6.70\gamma_{0}$ was used in place of the $6\gamma(d)$ in eqn (\ref{Es/dis}), corresponding to the cohesive energy for nearest and next nearest neighbors of a bulk filament in a hexagonal packing of spacing, $d$.}.
While the circular approximation of the faceted boundary shape for straight bundles leads to an underestimation of the surface energy change, we find that the continuum expression for $E_{side}$ effectively captures the shape and magnitude of surface energy decrease as the bundle is twisted.

\begin{figure}[h]
\centering
\includegraphics[width=8.0cm]{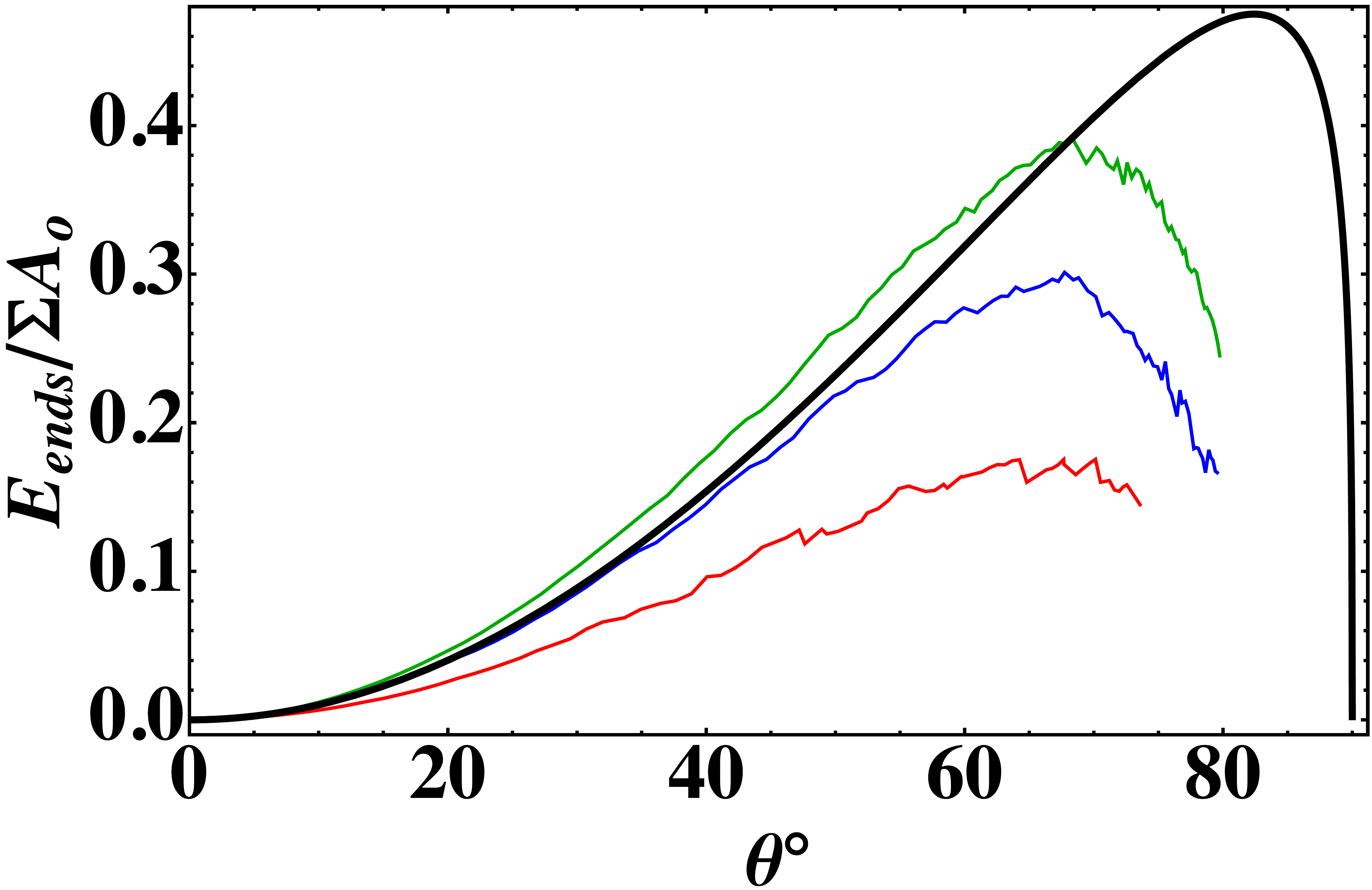}
\caption{End surface energy for three selected bundle sizes: small $N = 34$ (red), medium $N = 82$ (blue), and large $N = 184$ (green), in the infinite length limit. Black line is the continuum model expression eqn (\ref{eq:EndsLInfinity}).}
\label{fig:SurfaceEnergyDiscCont}
\end{figure}

Though not considered explicitly, the discrete model simulations of the previous section do implicitly count the cohesive energy loss due to radial slip at the filament boundaries, deriving from the curvature dependence of the cohesive energy. In the continuum limit where $\kappa_j d \ll 1$ we may approximate the curvature-dependent prefactor in eqn (\ref{discreteEij}) as
\begin{equation}
L\big( 1 + \kappa_j \pmb{\Delta}_{ij} \cdot {\bf N}_j \big)^{-1/2} \simeq L - \ell_s (ij) /2 ,
\end{equation}
where $\ell_s (ij) \simeq \kappa_j \pmb{\Delta}_{ij} \cdot {\bf N}_j = L_{ij}-L_{ji}$ is the difference in contacting length of $i$ with $j$ and contacting length of $j$ with $i$~\footnote{This identity derives from the mapping of curve $i$ to the point of contact on $j$, ${\bf R}_j(s_i) = {\bf R}_i(s_i)+ \pmb{\Delta}_{ij}(s_i)$. Since $d \pmb{\Delta}_{ij}/d s_j \cdot {\bf T}_j = -\kappa_j ({\bf N}_j \cdot \pmb{\Delta}_{ij})$, we have $|d {\bf R}_j/ ds_i| = \big(1+ \kappa_j \pmb{\Delta}_{ij} \cdot {\bf N_j}) \big)^{-1}$ and $L_{ji} \simeq L_{ij}\big(1- \kappa_j \pmb{\Delta}_{ij} \cdot {\bf N_j}) \big)$. }. Note that the implicit loss of contact from radial slip deriving from the curvature dependence does not account for the additional twist-dependent slip between azimuthally separated pairs (e.g.\ Fig.\ \ref{fig:SlipLenDia}b), which enters explicitly into $L_{ij}$. In the following section we generalize our discrete model to include these additional costs. We derive the surface contribution from loss of filament contact at the ends of twisted bundles of infinite length as 
\begin{equation}
E_{ends} (L \to \infty)/L =\sum_{ ij}  \gamma (\Delta_{ij}) \Big(\frac{1}{\sqrt{ 1 + \kappa_j \pmb{\Delta}_{ij} \cdot {\bf N}_j }} - 1 \Big),
\label{eq:EndsLInfinity}
\end{equation}
where again we take $i$ to be the outer filament of the pair so that $L_{ij} =L$. Fig.\ \ref{fig:SurfaceEnergyDiscCont} compares this surface energy formulation applied to our small, intermediate, and large discrete model bundles, to the continuum model prediction of eqn (\ref{Endlong}). This shows strong agreement over a large range of twist angles as the number of filaments per bundles grows sufficiently large.
As described in section \ref{sec:DiscreteEnergy}, we calculate the bulk cohesive energy shown in Fig.\ \ref{fig:InternalSimEn}, by subtracting the surface contributions given in eqns (\ref{Es/dis}) and (\ref{eq:EndsLInfinity}) the total, discrete model energy, that is, $E_{bulk} = E_{tot} - E_{sides} - 2 E_{ends}$.

\subsection{Thermodynamics of surface energy}
\label{sec:ContinuumThermo}
The continuum analysis in the previous section demonstrates that for sufficiently long bundles ($L\geq 5.07 R_{0}$) the cohesive energy at the surface prefers finite and generically large values of bundle twist. Along with the constraints and costs of packing frustration at the bundle core, the additional mechanical cost of {\it filament bending} competes with the surface energy preference for twist. In this section, we analyze an albeit simplified model of the twist dependence of cohesive bundles that includes only the {\it bending energy} and {\it surface energy}. Our goal is to determine the optimal twist geometry of filament bundles in the absence of the additional costs of filament packing in the bulk that were considered in section \ref{sec:CoreEnergy}, and which are considered again in the next section. The continuum analysis of these two competing energies suggests that minimal-energy bundles generically exhibit a degree of spontaneous twist that is highly sensitive to both bundle aspect ratio and filament stiffness.

The mechanical cost to bend a straight filament into a helical shape is simply $B \kappa^2 L/2$, where $B$ is the bending modulus, or stiffness, of the filament. In the continuum limit, we compute the total bending energy of the filaments in a twisted bundle by integrating over the cross-sectional area of the bundle
\begin{equation}
E_{bend} = \frac{B L}{2} \int dA \left(\frac{dN}{dA}\right) \kappa(\rho)^{2} , 
\label{eq:BendEnIntegral}
\end{equation}
where $dN/dA$ is the areal density of filaments in the {\it horizontal cross section} of the bundle. Following volume-conservation considerations similar to section \ref{sec:SurfGeometry}, assuming a volume fraction of filaments in the bulk of the bundle, the density at an area element located at $\rho$ is reduced by twist, according to,
\begin{equation}
\frac{dN}{dA} = \frac{n_{0}}{\sqrt{1+(\Omega \rho)^{2}}},
\label{eq:NumDensity}
\end{equation}
where $n_{0}^{-1} = (\sqrt{3}/2)d^2$ is the cross-sectional area per filament in a dense hexagonal packing. Combining both eqns (\ref{eq:BendEnIntegral}) and (\ref{eq:NumDensity}) to perform our integral in polar coordinates yields,
\begin{equation}
E_{bend} = \frac{ \pi B n_{0} L}{3} \left(2 - \frac{2+3(\Omega R)^{2}}{\left(1+(\Omega R)^{2}\right)^{3/2}}\right).
\label{eq:BendEnergy}
\end{equation}
Note that the form of $E_{bend}$ is not an explicit function of filament number or $R_{0}$, but instead depends only on twist angle $\theta = \arctan(\Omega R)$. In the limit of weakly twisted bundles, bending energy exhibits a soft dependence on twist $\lim_{\Omega R \ll 1} E_{bend} =(\pi B n_{0} L/4) (\Omega R)^4$, while in the limit of large twist, the bending cost asymptotically approaches a constant value $\lim_{\Omega R \gg 1} E_{bend} = (\pi B n_{0} L/3) \big[2 - 3/(\Omega R) \big]$

\begin{figure}[h]
\centering
\includegraphics[width=8.0cm]{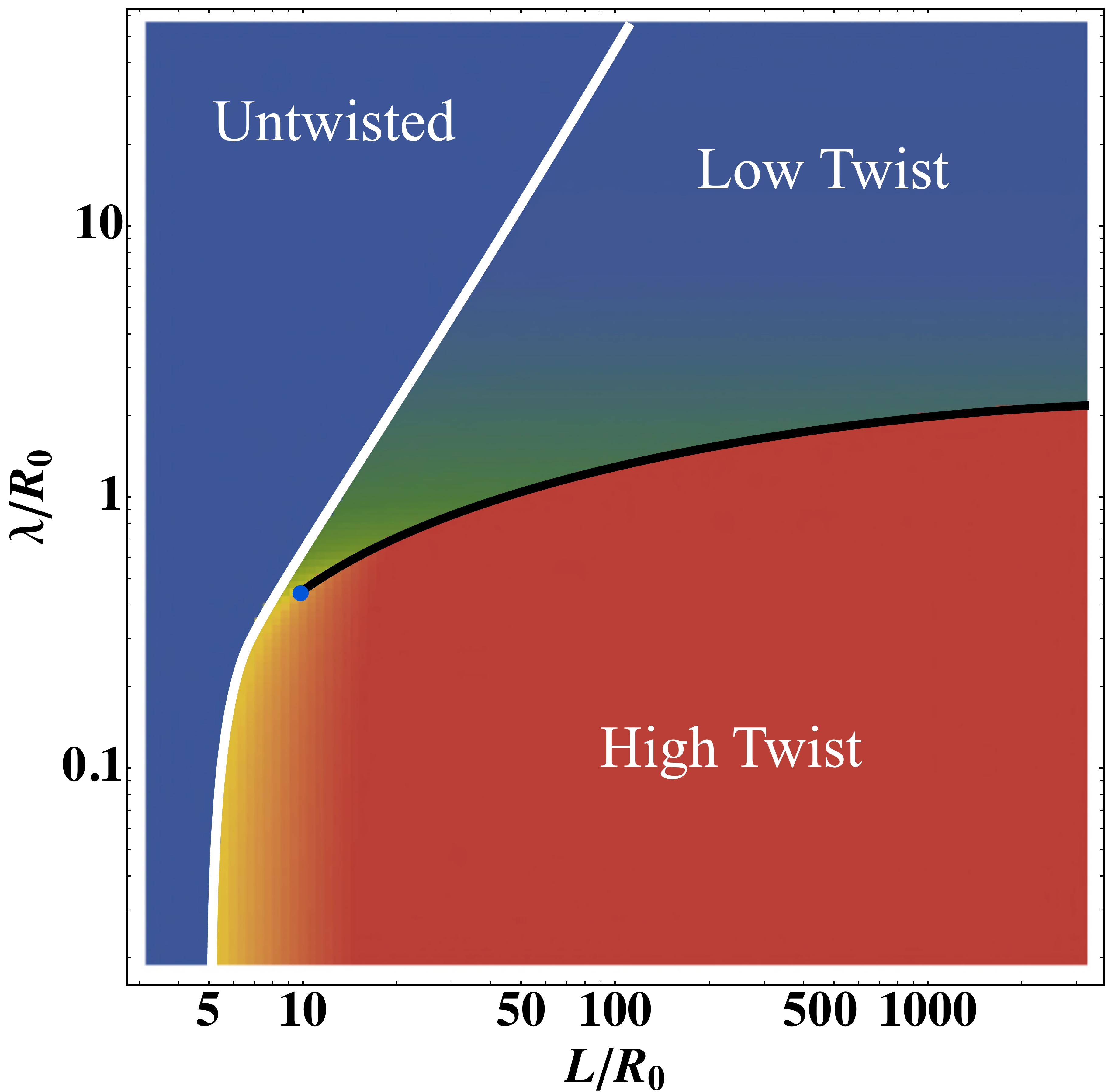}
\caption{Twist diagram of state of for the continuum energy, $E_{cont}$, showing three regions of optimal twist behavior: Untwisted (left of white line), bending energy dominated (upper right ``Low Twist'' blue region), and cohesive energy dominated (lower right ``High Twist''red region). The color represents the preferred value of $\theta$, ranging from $0^{\circ}$ (blue) to $90^{\circ}$ (red).}
\label{fig:BendLengthPhaseD}
\end{figure}

We define the total {\it continuum energy} as the sum of surface and bending energy contributions
\begin{equation}
E_{cont} = E_{surf} + E_{bend},
\label{eq:ContTotalEn}
\end{equation}
where again $E_{surf}= E_{side}+2E_{ends}$, as defined by the continuum expressions eqs. (\ref{eq:SideEn}) and (\ref{eq:EndSurfEn}). To compare the two types of energy, we define a length scale, 
\begin{equation}
\lambda = n_{0} B/ \Sigma_{0},
\end{equation}
which parametrizes the relative costs of bending to cohesive energies in filament assemblies. Optimizing $E_{cont}$ with respect to the twist angle for fixed $R_{0}$, $L$, and $\lambda$, we compute the diagram of state, shown in Fig.\ \ref{fig:BendLengthPhaseD}.

The phase diagram divides into three principle regions: untwisted ($L/R_0 \ll 1$), bending energy dominated ($\lambda/R_0 \gg 1;L/R_0 \gg 1 $), and cohesive energy dominated ($\lambda/R_0 \gg 1;L/R_0 \ll 1 $). I n the untwisted region, twisting a bundle is always unfavorable, due to the combined cost of slip at the ends of short bundles and bending. In the limit of highly flexible filaments, where $\lambda/R_{0} \ll 1$, the transition between untwisted and twisted bundles occurs at $L/R_{0} \simeq 5.07$; and as filament length grows, the balance of side and end surface energy yields an optimal twist that diverges with aspect ratio as $\Omega R \sim(L/R_{0})^{1/3}$. For larger stiffness, $\lambda/R_{0} \gg 1$, the more significant cost of filament bending shifts the boundary between untwisted and twisted states to larger aspect ratio as $L/R_{0}\sim (\lambda/R_{0})^{1/2}$. In this bending energy dominated region, there is a significant mechanical cost for bending filaments, however, for sufficiently large $L/R_{0}$, $E_{cont}$ is minimized by a modest twist of a few degrees. This arises from the fact that for small twist the cost of bending grows as $\sim B (\Omega R)^{4}$, while the surface energy decreases as $\sim \Sigma_{0} (\Omega R)^{2}$, leading to an optimal twist of $(\Omega R) \sim (R_{0}/\lambda)^{1/2}$. In this regime, the lowest energy state is nearly independent of the bundle length because the bending cost dominates the end effects of the surface energy term. Conversely, the amount of twist in the cohesively-dominated region ($L/R0\gg 1; \lambda/R_{0}) \ll1$) is largely only dependent on $L/R_{0}$. These two regions are separated by a first order transition for bundle lengths beyond a critical size $L/R_{0} \ge 9.9$. In the infinite length limit, the optimal twist angle jumps from $\theta \rightarrow 90^{\circ}$ to $14.0^{\circ}$, at $\lambda/R_{0} = 2.996$. For lower aspect ratios, this first order transition disappears, and the high and low twist energy minima merge into one. This critical point is shown as the blue dot in Fig.\ \ref{fig:BendLengthPhaseD} at $L/R_{0} = 9.9$, and $\lambda/R_{0} = 0.425$.

To summarize, we find that the balance of cohesive energy at the surface of sufficiently long bundles ($L> 5.07 R_{0}$) and flexibility favors large bundle twist. For bundles of rigid filaments, surface energy drives a more modest degree of spontaneous twists in minimal energy bundles above a critical aspect ratio that grows with filament stiffness.

\section{Optimal twist of ground-state bundles: Finite stiffness and length}
\label{sec:DiscreteFiniteL}
The previous sections have identified two competing effects of twisted filament bundle geometry: packing frustration of filaments in the bulk, and the surface energy cost of non-contacting filaments. In this section, we reexamine the energetics of our discrete filament bundle simulations, including the full costs associated with filament bending and loss of contact in bundles of finite length filaments. As described in section \ref{sec:DiscreteModel}, cohesive energy between filament pairs derives from the local contact geometry in twisted bundles. However, for the case of finite length filament bundles, the contact length of filament $i$ to $j$, $L_{ij}$ used in eqn (\ref{discreteEij}), must be calculated explicitly to account for {\it azimuthal slip} at the ends of the bundle (see Fig.\ \ref{fig:SlipLenDia}b). Defining the ends of filaments to be at positions $s_j \pm L/2$, $L_{ij}$ is calculated for any given length in terms of the contact function defined in section \ref{sec:DiscreteModel} as
\begin{equation}
L_{ij} = s_{i}(s_{j}^{*} = +L/2) - s_{i}(s_{j}^{*} = -L/2),
\label{eq:FiniteLCalc}
\end{equation}
where we follow our original notation that $s_{j}^{*}$ is the arc length coordinate of filament $j$ that is the point of contact with $s_{i}$. It can easily be shown that explicitly determining contact length for a filament pair along with the curvature-dependent correction to cohesive energy, properly accounts for the cost of both azimuthal and radial slip of filaments at the bundle ends.

\begin{figure}
\centering
\includegraphics[width=8.0cm]{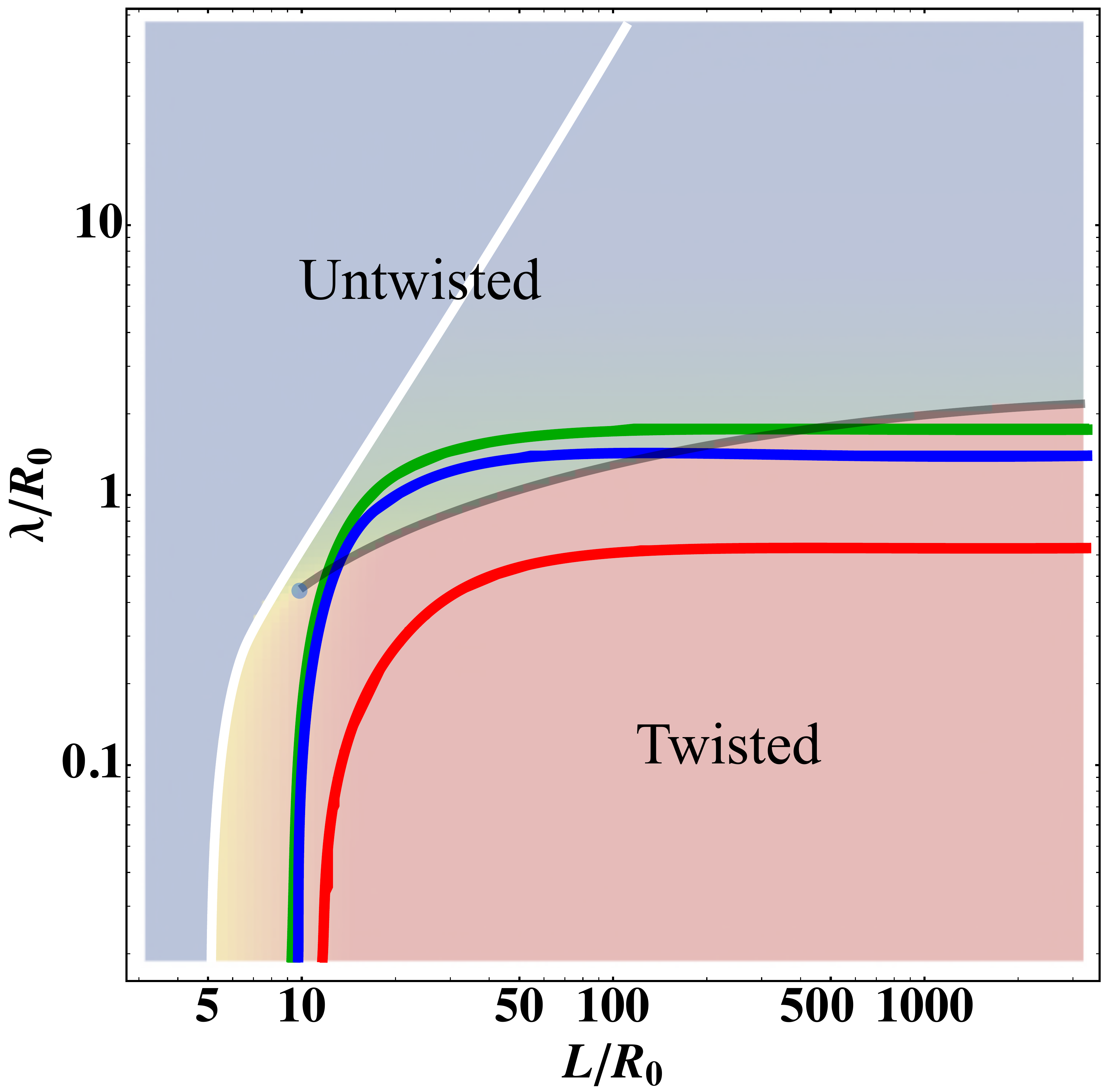}
\caption{Boundaries of the preferred state of filament bundles for three select sizes: small $N = 34$ (red), medium $N = 82$ (blue), and large $N = 184$ (green); overlaid on the continuum model results from Fig.\ \ref{fig:BendLengthPhaseD}. Above these boundaries bundles prefer an untwisted state, while below, the bundles can lower their energy by adopting a highly twisted state ($\theta \gtrsim 75^{\circ}$).}
\label{fig:SimEnPhaseD}
\end{figure}

Evaluating the total energy characterized by an aspect ratio $L/R_0$ and a finite stiffness corresponding to $\lambda/R_0$, we determine the optimal (energy-minimizing) value of twist. The phase boundaries separating untwisted and twisted ground states for small, intermediate, and large bundles are shown in Fig.\ \ref{fig:SimEnPhaseD}.
Importantly, we find that the cost of bulk packing, excluded from the continuum analysis of the previous section (see Fig.\ \ref{fig:BendLengthPhaseD}), in combination with the bending cost, eliminates the regime of stable, weakly-twisted bundles in the bend-dominated region, $\lambda/R_{0} \gg1$, that is predicted by in the continuum model. The discrete model of cohesive filament bundles exhibits only two well-defined phases: untwisted bundles for sufficiently short or rigid filaments, or highly twisted with a twist angle of $\theta \gtrsim 75^{\circ}$ for long and flexible filaments. The bundles in this highly twisted state have the maximum topological charge of $Q = 6$.

In the limit of infinitely flexible filaments ($\lambda / R_{0} \ll 1$), we find that the inclusion of bulk energy shifts the predicted critical aspect ratio required for twist from $L/R_{0} \ge 5.07$, to $L/R_{0} \gtrsim 8.9$. In the other limit of infinitely long filaments ($L/R_{0} \gg 1$), the critical value of $\lambda / R_{0}$, above which the filaments are stiff enough to resist the surface energy drive to twist, is $\lambda / R_{0} \ge 0.63$, $\lambda / R_{0} \ge 1.41$, and $\lambda / R_{0} \ge 1.78$, for $N=34$, $N=82$, and $N=184$, respectively.

These results show that the additional costs of packing frustration in the bulk significantly offsets the gain in cohesive energy derived from the surface of long bundles. We find that increasing the number of filaments generically increases the range over which minimal energy bundles are twisted, substantially raising the threshold stiffness for the boundary between twisted and untwisted states. As a final comment, we note the appearance of highly-twisted ground states of our discrete simulation model, above the first order line separating highly-twisted from weakly-twisted bundles in the oversimplified continuum model predicted by optimizing $E_{cont}$ alone, which derives specifically from the underestimation of $E_{side}$ (and the driving force for twist) in the continuum approximation (see Fig.\ \ref{fig:BendLengthPhaseD}).

\section{Discussion}
\label{sec:Discussion}
We conclude with a discussion of our results, highlighting two specific aspects. First, we summarize the role of excess disclinations as key structural elements to the ground-state order of twisted bundles, and discuss the implications of kinetic pathways of assembly for twisted bundles. Second, we review the predictions of our model with respect to three specific systems of cohesive filament assemblies.

\subsection{Defects and kinetic limitations to ground-state packing}
\label{sec:Kinetic}
While the preference for twist is driven by effects at the boundary of the bundle, the complex evolution of cross-sectional packing---as evidenced by the universal increase in topological charge of the packing---plays a critical thermodynamic role in stabilizing twisted bundles. Above a critical threshold of twist $\theta \simeq 22^\circ$, excess 5-fold disclinations are needed in the ground-state packing to {\it screen} the elastic effects of the packing frustration generated by twist. Continuum elasticity arguments have show that twist decreases inter-filament spacing between azimuthally-separated neighbors by an amount proportional to $(\Omega \rho)^2$, ultimately leading to an increase in energy (per unit volume) that grows as $(\Omega R)^4$ for defect-free bundles \cite{Grason2012}. Hence, in the absence of topological defects which act to ``neutralize" the stresses generated by twist, the elastic cost of twisting defect-free bundles would continue to grow unmitigated at large twist angle, likely overwhelming the gains in cohesive energy at the boundary.

\begin{figure}[h]
\centering
\includegraphics[width=8.0cm]{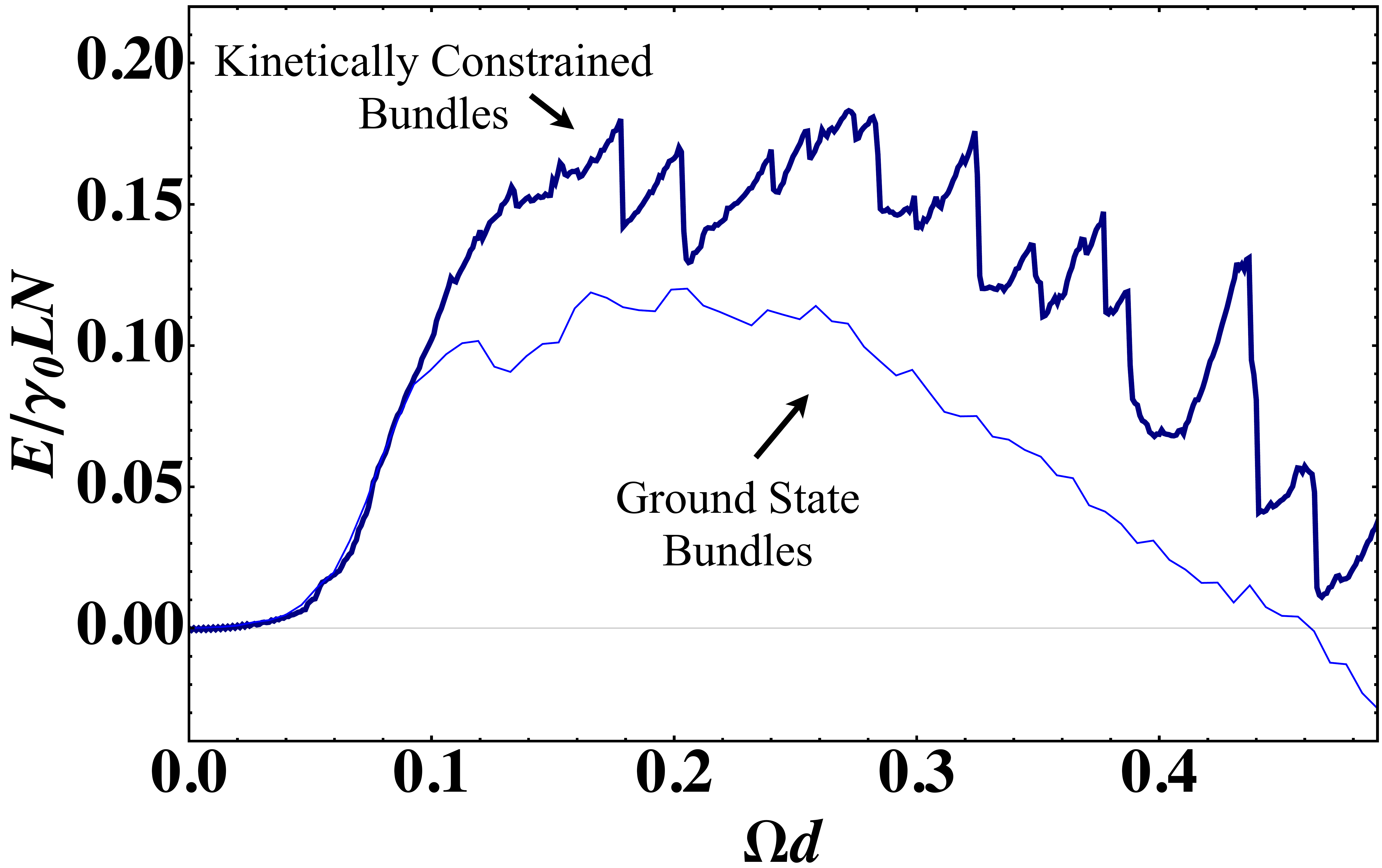}
\caption{Lower thin line shows the thermodynamically lowest energy ground states vs twist rate $\Omega d$ (same as shown in Fig.\ \ref{fig:TotalSimEn}). Upper thick lines represent the constrained ground state energies. $N=82$ for both.}
\label{fig:KineticTrap}
\end{figure}

We demonstrate the importance of achieving the appropriate defect configuration for stabilizing twist by considering a class of {\it kinetically-constrained} bundle packings in our numerical simulation model. Unlike our numerical search for ground states described in section \ref{sec:CoreEnergy}, which explored an ensemble of in-plane packings at each value of twist, in Fig.\ \ref{fig:KineticTrap} we analyze the energetics of simulated packings achieved in the following kinetically-constrained algorithm. Beginning from the energy-minimized packing of an untwisted bundle, we increase $\Omega$ in small increments of $0,001 d$. For each $\Omega$, we perform a steepest-descent minimization of bundle energy (in the $L \to \infty$ limit) based on the starting positions of the previous, smaller value of twist.
Fig.\ \ref{fig:KineticTrap} compares the energy of kinetically-constrained bundles to the simulated ground states presented in section \ref{sec:DiscreteEnergy}. For small $\Omega d$, the cohesive energies of both states are identical as expected since no large-scale filament reorganizations are required. This persists until a high enough twist forces the ground state bundle into a new configuration at $\Omega d \approx 0.1$. At this point they cannot reach this new ground state packing as it requires a global rearrangement of the filaments. For example, as shown Figs.\ \ref{fig:BundleSpread} and \ref{fig:SimPhaseD}, at the transition from $Q=0$ to $Q=1$ packings the minimal-energy packing changes from a defect-free hexagonal packing to packing possessing an excess 5-fold defect near the bundle center. In the kinetically-constrained packings, such defects only enter at the boundaries of the bundle, migrating only slowly towards the center upon further twisting. Fig.\ \ref{fig:TotalSimEn} shows that further increase in twist eventually does allow the constrained bundle to overcome the local energy barriers of filament arrangement; however, the total energy of states continues to exceed the ground state packings inhibiting the stability of the twisted bundle relative to the $\Omega =0$ state.

The discrepancy between ground state packings and this simple model of kinetically-constrained bundles demonstrates two key points about disclinations in twisted filament bundles. First, changes in surface energy will only stabilize twisted bundles over straight bundles provided that the appropriate, energy-minimizing configuration of disclinations punctuates the cross-sectional packing. And second, the kinetic pathways by which topological defects enter into and migrate throughout the cross section may place strong constraints on whether an externally or intrinsically twisting system of filaments is able to achieve an optimally-twisted state. This points to the need for further work on the mobility of disclinations and the plasticity in twisted filament arrays.

\subsection{Implications for cohesive filaments assemblies}
\label{sec:RealSys}
We now consider how the predictions of our model apply to cohesive assemblies of filaments from a range of systems in materials and biology, whose properties vary in terms of size, stiffness and cohesive interactions. While we opted to use a Lennard-Jones potential to describe interfilament forces in our discrete model, we find ultimately that the predominant thermodynamic sensitivity of bundles to twist depends only on two primary quantities characterizing the interaction: $\gamma_{0}$, the depth of the cohesive interactions (per unit length); and $d$, the preferred local spacing between bound neighbors. Thus, it is natural to extend the predictions of the current study to filament systems whose finite-range cohesive forces are not explicitly modeled by our ``Lennard-Jones thread" model, provided appropriate values of $\gamma_0$ and $d$~\footnote{In principle, the relative cost of the surface exposure to twist-induced frustration is also sensitive to the ``stiffness", or second derivative, of the potential, which controls the elastic properties of the array. The stiffness of the ``LJ thread" potential in eqn (\ref{gamma}), is of order $\sim \gamma_0/d^2$, hence, we expect any system interacting via a similarly soft potential to be well described by the large-$N$ results of the present model.}. Indeed, it is a key finding of the present study that the ground-state packing, characterized in terms of topological charge $Q$, is entirely insensitive to even these features of the inter-filament potential, either its depth {\it or} preferred separation.

Assuming that the assembly kinetics of bundles accommodate the appropriate number and distribution of disclinations, we find that regimes of thermodynamically preferred bundle twist are separated from untwisted bundles by characteristic measures of bundle size. As shown in section \ref{sec:DiscreteFiniteL}, our discrete model calculations suggest that equilibrium bundles are spontaneously twisted when $L \gtrsim 10 R_0$ and $R_0 \gtrsim \lambda$. Unlike the aspect ratio, which is a purely geometric parameter, $\lambda$ varies with intrinsic properties of filament stiffness and cohesive forces. We present a brief consideration of the value of $\lambda$ for three distinct filamentous systems, with the goal of assessing the thermodynamic stability of each to bundle twist: (i) capillary-condensed arrays of micro- and nano-fabricated pillars; (ii) carbon nanotube ropes; and (iii) DNA bundles condensed in the presence of polyvalent counterions. We quantify cohesive tendency for twist in terms of $N_c \equiv (\lambda/d)^2$, roughly the number of filaments needed to stabilize surface-driven twist.

{\it (i) Capillary condensed filaments} - On the upper end of filament diameters, we find filament arrays held together by capillary forces, as occurs when filament arrays are drawn from a wetting into to non-wetting fluid medium \cite{Roman2010}. In such cases, inter-filament cohesion is mediated by liquid bridges spanning neighboring filaments in the array. As such, we expect the surface energy of the bundle, $\Sigma_0$, to be proportional to the surface tension between wetting and non-wetting fluids, of order $\sim 10$ mN/m$^2$ ~\cite{israelachvili2011intermolecular}. Since stiffness is a strong function of filament diameter $B \sim d^4$, bundles of large diameter filaments, such as hair \cite{Bico2004} and similarly sized glass or polystyrene fibers \cite{Py2007}, with diameters $d \gtrsim 10~{\rm \mu m}$ and bending stiffness in the range $B \sim 3~{\rm mN~mm^{2}}$ to $3000~{\rm mN~mm^{2}}$, are relatively stable to twist, only twisting for filament numbers larger than $N_c \approx$ 3000. On the other hand, arrays of more slender nano-fabricated pillars $d \approx 300~{\rm nm}$ are relatively easy to twist by capillary forces even for bundles of just a few filaments, as $N_c \approx 1$, consistent with observations of ref.\ \cite{Pokroy2009}.

{\it (ii) Nanotube ropes} - On the opposite end of the size spectrum are ropes of single-walled carbon nanotubes, with diameters typically in the range of $d \approx 8$~\AA~(for (6,6) SWNTs) and $d \approx 27$~\AA~(for (20,20) SWNTs). Ropes of carbon nanotubes are prepared by a variety of methods, from the electric-arc discharge of graphite \cite{Journet1997} to nano-textiles spun from grown nanotube mats \cite{Zhang2004}. Nanotube ropes are typically formed in the limit of extreme aspect ratio, $L/R_{0} \gg 100$. While the influences of tube chirality, metallic/semi-conducting and polydispersity properties of nanotubes complicates the simplistic treatment of inter-filament cohesion considered here, twisted structures have been reported in bundles containing at least tens of the SWNT \cite{Henrard2000a, Zhang2002}. Depending on the nanotube diameter estimates for stiffness vary considerably, ranging from $B \approx 30~{\rm nN ~nm^{2}}$ for (6,6) tubes to $4000~{\rm nN~nm^{2}}$ for (20,20) tubes, while van der Walls attraction between nanotubes in vacuum suggestion a cohesive energy per length of $\gamma_0 \approx800 ~ {\rm pN}$ relatively independent of diameter ~\cite{Liang2005,Girifalco2000}, from which we estimate $\lambda \approx 80 ~{\rm nm}$ and $2 ~{\rm \mu m}$ for small and large diameter tubes, respectively. The large value of $\lambda/d$ implies that nanotubes are fairly rigid despite their small diameter, presumably due to intrinsic stiffness afforded by covalent bonding within tubes. These estimates suggest a very modest tendency for nanotube ropes to twist, which varies considerably with tube diameter: $N_{c}$ of 5,000 and 500,000 nanotubes for single tube diameters of $ 8$~\AA~ and $27$~\AA~ respectively.

{\it (iii) Condensed DNA bundles} - dsDNA condenses in solutions of multivalent counterions into tightly packed toroids \cite{Hud2001, Conwell2003, Leforestier2009} and bundles (sometimes referred to in the literature as ``rods'')\cite{BIP:BIP360300514, Leforestier2009, McLoughlin2005}. Given a bending rigidity of $B \approx 0.24~{\rm nN~nm^{2}}$ ~\cite{sinden1994dna, Geggier2010}, an interaction energy per unit length of $\gamma_0 \approx 6 ~ {\rm pN}$, and a center-to-center spacing of condensed DNA, $d \approx 3~{\rm nm}$~\cite{Qiu2011a} in the presence of trivalent cations, we can estimate $\lambda \approx 13~{\rm nm}$. This sets a critical number of cross-sectional DNA strands to stabilize twist as $N_c \approx 18$. This result implies that the relative flexibility of dsDNa (in comparison to, say, carbon nanotubes) is overwhelmed by inter-strand cohesion in nominally sized-bundles, and cohesive forces alone may be sufficient to stabilize twist in toroidal bundles of dimensions typical for encapsulated bacteriophage genomes \cite{Purohit2005, Leforestier2009, Petrov2008, Leforestier2011}.

The simple model estimates above neglect many key aspects of inter-filament forces that may further stabilize or inhibit twist in cohesive bundles. Notably, the present model does not account for the twist dependence of interactions between chiral filaments, a feature well-known to bias the handedness and drive the twist of interfilament packings in condensed phases of helical molecules from DNA to collagen \cite{Kornyshev2007, Harris1999}. Surprisingly, the broad range of filament sizes and (achiral) cohesive forces considered here suggest that even in the absence of intrinsic or external driving torques, thermodynamically preferred twist is the {\it rule rather than the exception} in cohesive bundles of long and flexible filaments. Furthermore, this feature persists despite the inclusion of defects within the cross-sectional packing of sufficiently twisted bundles.

\section{Conclusion}
Based on a detailed analysis of the discrete numerical model and the continuum surface model for the cohesive energy in twisted bundles, we find that the lowest energy state for a bundle of sufficiently flexible and long filaments is generically twisted. The decrease in energy with twist derives primarily from a decrease in the surface energy of the bundles, which ultimately accounts for the fact that twisted bundles expose {\it fewer} filaments at the high-energy sides than straight bundles.

The cohesive assemblies of molecular filaments studied here share a common structural design and are in fact quite similar to non-cohesive macroscale twisted filament bundles such as yarns, ropes, and cables. While the present mode ignores important features of macroscale materials such as inter-filament friction\cite{hearle1969structural}, it does however shed light on the generic and complex relationship between the defects necessary for twist and the internal mechanics of a bundle. While the low energy packing structures studied in the present model form in the absence of external loads, it is clear that the global application of external stress to the bundle will modify the mechanical driving forces that favor lattice defects. For instance, when filaments in a twisted bundle are under tension (say, when the bundle is stretched), outer filaments exert radial compressive stresses on the internal filaments, whose magnitude varies with radial depth. Hence, it is natural to anticipate that applied tension may become an additional axis to the phase diagrams presented in Figs.\ \ref{fig:SimPhaseD}, \ref{fig:BendLengthPhaseD} and \ref{fig:SimEnPhaseD}. We speculate that the low-energy states of twisted bundles will be highly sensitive to the presence of external stress and may vary dramatically in structure from bundles forming in the absence of externally imposed mechanical forces. Furthermore, since the presence of lattice defects considerably alters the internal stresses of twisted bundles, it remains an outstanding and important challenge to understand how the presence of energy-minimizing defects in cohesive filament bundles modify their emergent mechanical properties such as compliance, strength and flexibility.

\section*{Acknowledgments}
The authors are grateful to A. Parsegian for helpful discussions. This work was supported by NSF CAREER Award DMR 09-55760, the UMass Center for Hierarchical Manufacturing, NSF CMMI 10-25020, and an award from the Alfred P. Sloan Foundation.

\footnotesize{
\bibliography{rsc} 
\bibliographystyle{rsc} 
}

\begin{appendix}
\section{Bundle-equivalent surface}
\label{sec:DomeDerivation}
\begin{figure}[h]
\centering
\includegraphics[width=8.0cm]{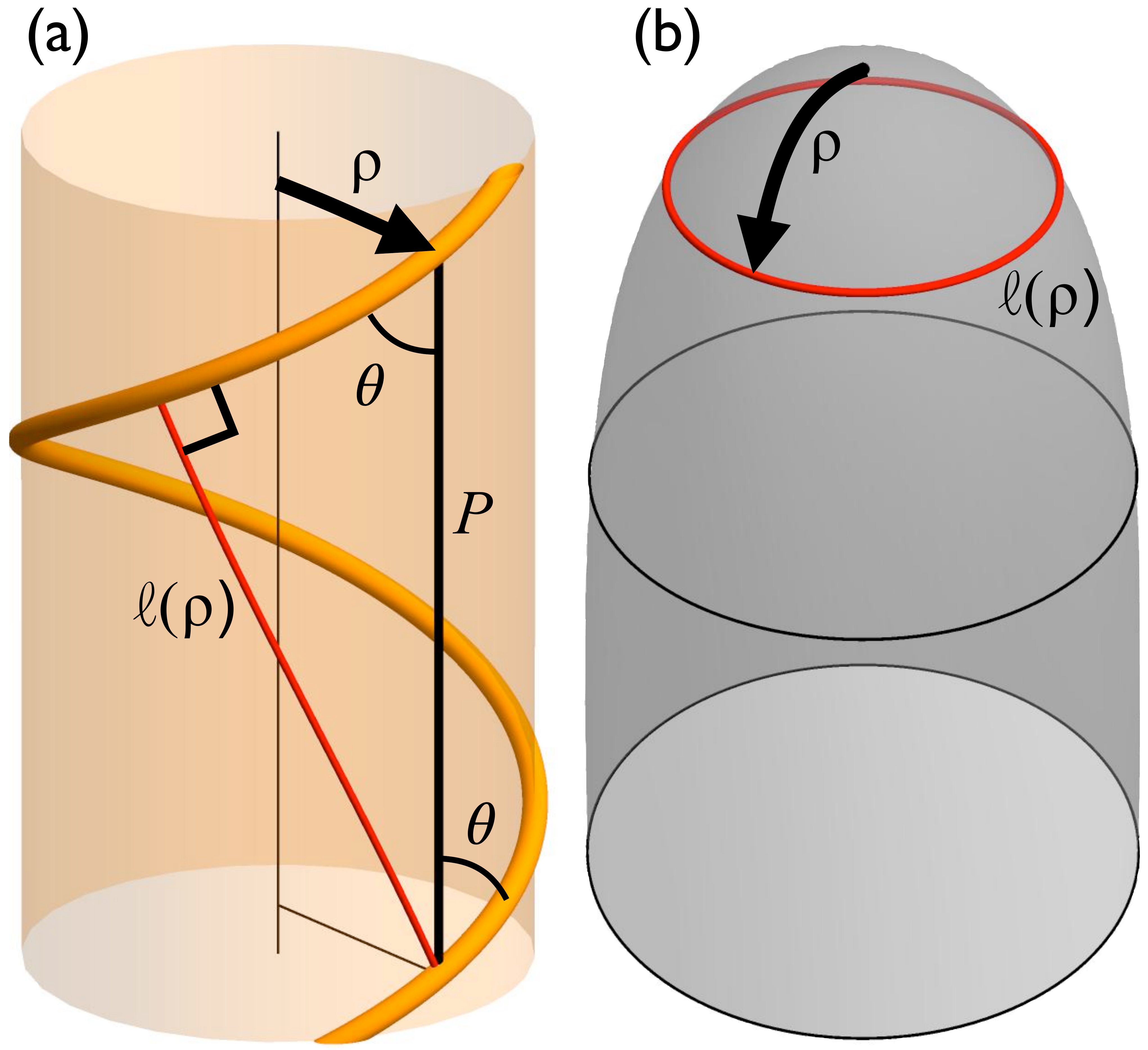}
\caption{(a) A helical filament (orange tube). (b) The bundle-equivalent surface.}
\label{fig:DomePara}
\end{figure}

The geometry of the {\it bundle-equivalent surface} can be constructed from simple considerations of the space available for packing filaments at a radial distance $\rho$ from the center of a twisted bundle, characterized the length $\ell (\rho)$ of a span between points of ``self-contact" on a helical curve. In terms of the dual surface representation, $\ell (\rho)$ corresponds to the circumference of the surface an arc-distance $\rho$ from its pole. This length is determined by considering a helical curve at $\rho$ and the length of the shortest, constant-radius span that connects the curve to itself at another point along its distance. The finite length of the span between ``self-contacts" of a helical filament at $\rho$ implies that the number of finite-diameter filaments that may be packed at a given radius is limited, as is the space available for packing finite-diameter discs at a given radius from the center of an axi-symmetric surface. Fig.\ \ref{fig:DomePara} shows the geometry of this span whose length derives from the helix geometry,
\begin{equation}
\ell(\rho) = P \sin \theta(\rho) = \frac{ 2 \pi \rho}{ \sqrt{ 1+ (\Omega \rho)^2 } } .
\end{equation}
The circumference of the surface grows with $\rho$ more slowly than $2 \pi \rho$, imply a spherically-curved geometry, characterized by a positive Gaussian curvature,
\begin{equation}
\label{KG}
K_G(\rho) = \frac{ 3 \Omega^2}{ \big[ 1 + (\Omega \rho)^2 \big]^2} ,
\end{equation}
showing that curvature (and hence geometric frustration) is concentrated at the pole of that surface (i.e.\ $\Omega \rho \ll 1$), corresponding to a region near the center of twist in a bundle.

We now consider the relationship between the geometry of the dual surface and the topology of the triangular network of nearest-neighbor bonds of discs or particles packed on this surface (see e.g. Fig.\ \ref{fig:DomePara}b), whose constraints carry over to the packing geometry of twisted bundles. For a triangular element, connecting three vertices of the packing, the Gauss-Bonnet theorem relates the integrated Gaussian curvature of the surface {\it patch} within the element to the {\it internal} angles, $\theta_i$, at vertices,
\begin{equation}
\int_{\rm patch} dA ~K_G + \pi = \sum_{i=1}^3 \theta_i,
\end{equation}
where we have taken the edges of the patch to be geodesics. Summing this over the entire mesh of the packing, we have
\begin{equation}
\int_{\rm mesh} dA~ K_G + \pi F = 2 \pi \sum_n V(n) + \sum_{b} \theta_b ,
\label{GB2a}
\end{equation}
where $F$ is the total number of faces in the mesh, $V(n)$ is the number of $n$-fold vertices not at the boundary of the triangulation (each of contributing $2 \pi$ from the sum of interior angles), and $\theta_b$ are the interior angles of vertices on the {\it boundary} of the triangulation. Defining $V_b(n)$ as the number of $n$-fold vertices at the boundary and using $3 F= \sum_n \big[ n V (n) + (n-1) V_b (n) \big]$, we may rewrite eqn (\ref{GB2a}) as
\begin{equation}
\int_{\rm mesh} dA ~K_G  = 2 \pi \sum_n \Big( 1 - \frac{n}{6} \Big)V(n) + \sum_{b} \Big( \theta_b - \frac{ \pi}{3} \Big) ,
\label{GB2}
\end{equation}
where we note that an $n$-fold boundary vertex possesses $(n-1)$ interior angles. Dividing by $2 \pi$ and making use of the definition of topological charge in eqn (\ref{eq:DiscCharge}), we have
\begin{equation}
\label{euler}
6 \chi  - Q = \frac{ 1}{2 \pi} \sum_b \Big( \theta_b - \frac{ \pi}{3} \Big) ,
\end{equation}
where we have defined
\begin{equation}
\chi  = \frac{1}{2 \pi} \int_{\rm mesh}  dA ~K_G,
\end{equation}
which plays the role of the Euler characteristic for a boundary-free surface domain, notably increasing as the lateral size of the patch grows large in comparison the curvature radius of the surface (proportional to $P$). The right-hand side of eqn (\ref{euler}) represents distortion of the nearest-neighbor packing from an equilateral geometry ($\theta_b \neq \pi /3$) at the free boundary of cluster, such that the deficit between the topological charge of the interior packing and $6 \chi$ must be distributed as boundary distortion of the packing. Approximating the boundary geometry of the packing as circular and integrating the Gaussian curvature within a packing of arc-radius $R$ we find,
\begin{equation}
\chi( R) = 1 - \frac{1}{ [ 1+ (\Omega R)^2 ]^{3/2}} ,
\end{equation}
which increases from 0 at small twists as $ 3 (\Omega R)^2 /2$, to a maximum of $1$ in the limit $|\Omega R| \to \infty$. By defining $Q_{id} = 6 \chi(R)$ as the ideal disclination charge that perfectly neutralizes the integrated Gaussian curvature in eqn (\ref{euler}), we arrive at eqn (\ref{eq:QidSim}) in the main text. Specifically, when the actual topological charge can achieve the ideal value ($Q = Q_{id}$), the distortion of the packing at the outer boundary from equilateral geometry ($\theta_b = \pi/3$) can vanish.

\section{Lattice orientational dependence of surface energy}
\label{sec:OrentationAlpha}
\begin{figure}[h]
\centering
\includegraphics[width=8.0cm]{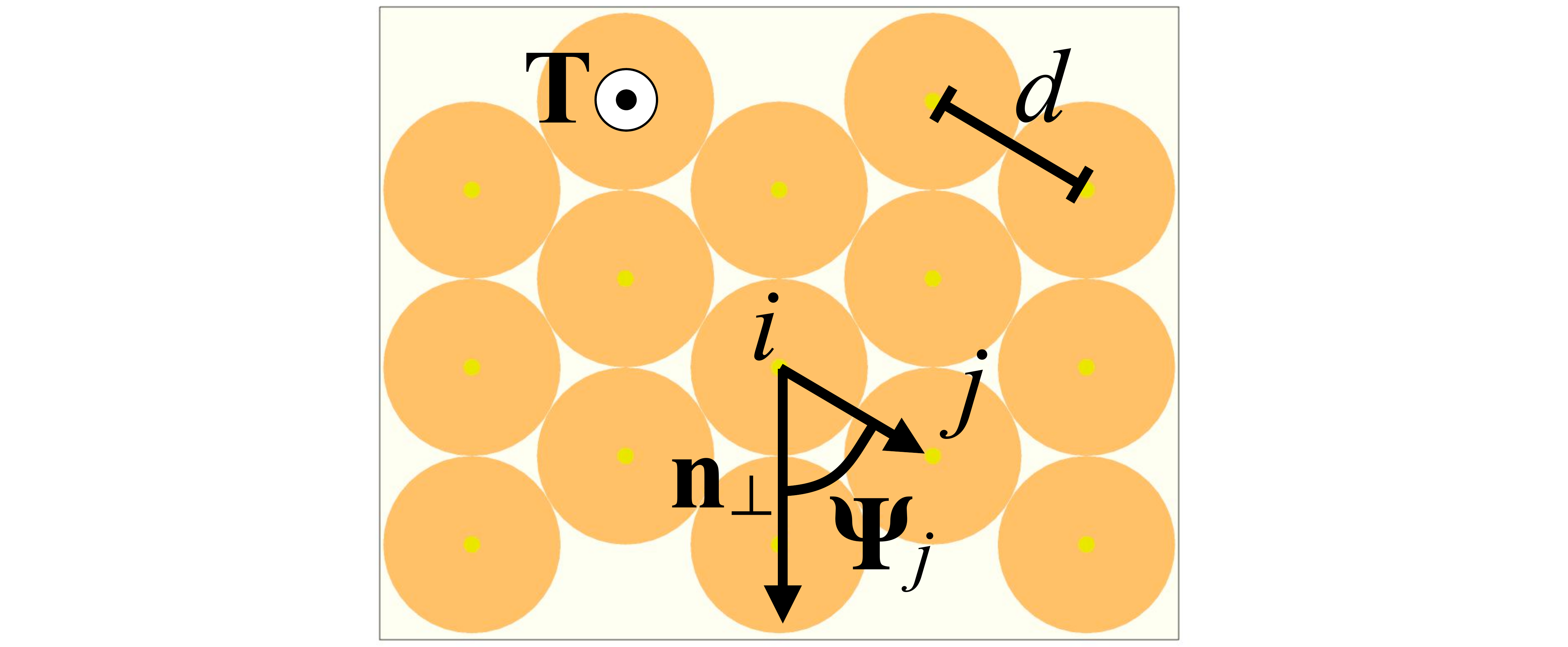}
\caption{Top view of bulk filaments in Fig.\ \ref{fig:BulkCuts} cut by a plane with the normal, $\bf{n}$. Filament $i$ and its neighbor $j$ are specified. The interfilament spacing is $d$, and $\Psi$ is the angle between the horizontal cross section component of the normal, $\bf{n}_{\perp}$, and the vector $\bf{X}_{j} - \bf{X}_{i}$.}
\label{fig:TopSurfCut}
\end{figure}

As described by eqn (\ref{eq:SurfEnergy}), the energy per unit of exposed surface area of a bundle, $\Sigma$, is dependent on the angle, $\Theta$, between the local filament tangent, $\bf{T}$, and the cutting plane normal vector, $\bf{n}$. This definition separates the surface corresponding only to the loss in contact lengths between filament pairs, as opposed to surface area associated with filament {\it ends}. However, as noted in section \ref{sec:ContinuumSurfEn}), there is an additional dependence of $\Sigma$ on the orientation of the cutting plane with respect to the lattice directions at the surface, which is defined in terms of the component of $\bf{n}$ that lies in the horizontal cross section, (shown in Fig.\ \ref{fig:TopSurfCut})
\begin{equation}
\bf{n}_{\perp}= \frac{{\bf n} - {\bf T} ({\bf n}\cdot {\bf T} )}{ 1- ({\bf n}\cdot {\bf T} )^2 } .
\end{equation}

We consider the loss of contact for a single filament, $i$, for each its six nearest neighbors, $j$,
\begin{equation}
\ell_s(j) =  |\tan\Theta| d |\cos \psi_j|,
\end{equation}
where $\psi_j$ is the angle between $\Delta_{ij}$ and ${\bf n}_{\perp}$. Defining the angle of this orientation as $\Psi$, the {\it smallest} angle between the bond directions in the hexagonal lattice and ${\bf n}_{\perp}$ we have $\psi_j = \Psi+\pi j / 3$. Using the fact that area per filament at the cutting surface is $n_0^{-1} \sec \Theta$ and defining the dimensionless parameter $\alpha$ as in eqn (\ref{eq:SurfEnergy}), we have
\begin{equation}
\alpha(\Psi) = \frac{ n_{0} d^2}{2}  \sum^{5}_{j=0}\big|\cos\big( \Psi + \pi j /3\big) \big|  ,
\label{eq:TotalSlipSum}
\end{equation}
where $n_{0} = 2/\sqrt{3}d^{2}$. Summing over the filament neighbors we have,
\begin{equation}
\alpha(\Psi) = \frac{ 4}{\sqrt{3} } \cos \Psi , \ {\rm for } \ - \pi/6 < \Psi <\pi/6 .
\end{equation}
Two limiting cases of $\alpha$: 1) the lattice vector is aligned with $\bf{n}_\perp$ (high surface energy), yielding a maximum $\alpha(\Psi=0) = 4/\sqrt{3}$; and 2) the lattice vector is $\Psi = \pm \pi/6$ maximally offset from $\bf{n}_\perp$ (low surface energy), yielding a minimum $\alpha(\Psi=\pi/6) = 2$. For surface elements at the ends of the bundle, the distribution of $\Psi$ roughly visits all orientation equally, hence, suggesting the appropriate value of $\alpha$ is average with respect to $\Psi$:  $\langle\alpha\rangle = 4 \sqrt{3}/\pi \simeq 2.2$.

\end{appendix}
\end{document}